\documentclass[12pt]{article}
\usepackage{amsfonts}
\usepackage{amsmath,amssymb,amsthm}
\usepackage{appendix}
\usepackage{bbm} 
\usepackage{amsbsy}
\usepackage{enumerate}
\usepackage{cite}

\def\macrosH{}
\def\macrosHarxiv

\InputIfFileExists{./macros_local.tex}{}{}

\ifdefined\macrosPa
  \usepackage[textwidth=465pt,textheight=650pt,centering]{geometry} 
\else\ifdefined\macrosPb
  \usepackage[textwidth=500pt,textheight=650pt,centering]{geometry} 
\fi\fi

\ifdefined\macrosS


  \usepackage{mathptmx}
  \DeclareMathAlphabet{\mathcal}{OMS}{cmsy}{m}{n}
\fi

\ifdefined\macrosBirk
\else
\usepackage[dvips]{graphicx}
\fi

\ifdefined\macrosSB

\def\UseSection{
        \numberwithin{equation}{section}
	\theoremstyle{plain}
        \newtheorem{theorem}    {Theorem}[section]
        \DefineTheorems 
}

\def\DefineTheorems{
	
	\newtheorem{lemma}      [theorem] {Lemma}
	
	\newtheorem{prop}       [theorem] {Proposition}
	
	\newtheorem{cor}        [theorem] {Corollary}

	\theoremstyle{definition}
	\newtheorem{defn}       [theorem] {Definition}
	
	\newtheorem{example}       [theorem] {Example}

	\theoremstyle{definition}

}

\newcommand{\bt}   {\begin{theorem}}
\newcommand{\et}   {\end  {theorem}}
\newcommand{\bl}   {\begin{lemma}}
\newcommand{\el}   {\end  {lemma}}
\newcommand{\bp}   {\begin{prop}}
\newcommand{\ep}   {\end  {prop}}
\newcommand{\bc}   {\begin{cor}}
\newcommand{\ec}   {\end  {cor}}
\newcommand{\bd}   {\begin{defn}}
\newcommand{\ed}   {\end  {defn}}

\newcommand{\ba}   {\begin{array}}
\newcommand{\ea}   {\end  {array}}
\newcommand{\be}   {\begin{enumerate}}
\newcommand{\ee}   {\end  {enumerate}}
\newcommand{\bi}   {\begin{itemize}}
\newcommand{\ei}   {\end  {itemize}}

\def\eq#1\en{\begin{equation}#1\end{equation}}  
\def\eqsplit#1\ensplit{
	\begin{equation}\begin{split}#1\end{split}\end{equation}
	}
\def\eqalign#1\enalign{
	\begin{align}#1\end{align}
	}
\def\eqmul#1\enmul{
	\begin{multline}#1\end{multline}
	}
\newcommand{\eqarrstar} {\begin{eqnarray*}} 
\newcommand{\enarrstar} {\end{eqnarray*}} 
\newcommand{\eqarray}   {\begin{eqnarray}} 
\newcommand{\enarray}   {\end{eqnarray}} 
\newcommand{\nnb}	{\nonumber \\} 

\newcommand{\lbeq}[1]  {\label{e:#1}}
\newcommand{\refeq}[1] {\eqref{e:#1}}    

\makeatletter
\newcommand{\labelcounter}[2]{{%
	\stepcounter{#1}
	\protected@write\@auxout{}%
	{\string\newlabel{#2}{{\csname the#1\endcsname}{\thepage}}}%
	{\ref{#2}}
	}}
\makeatother
\newcommand{\Nbold} {{\mathbb N}}

\newcommand{\Rbold} {{\mathbb R}}

\newcommand{\Zbold} {{\mathbb Z}}

\newcommand{\Ncal}   {\mathcal{N}}

\newcommand{\spose}[1] {{\hbox to 0pt{#1\hss}} }
\newcommand{\ltapprox} {\mathrel{\spose{\lower 3pt\hbox{$\mathchar"218$}}
 \raise 2.0pt\hbox{$\mathchar"13C$}}}
\newcommand{\gtapprox} {\mathrel{\spose{\lower 3pt\hbox{$\mathchar"218$}}
 \raise 2.0pt\hbox{$\mathchar"13E$}}}

\else
\fi

\UseSection   
\usepackage[usenames]{color}

\definecolor{bw}{RGB}{240, 120, 0}
\definecolor{at}{rgb}{0.0, 0.5, 0.0} 

\renewcommand{\to} {\rightarrow}

\newcommand{\R}{\Rbold}
\newcommand{\Z}{\Zbold}

\newcommand{\N}{\Nbold}
\newcommand{\C}{\mathbb{C}}

\newcommand{\1}{\mathbbm{1}}

\newcommand{\psib}{\bar\psi}

\newcommand{\zetab}{\bar{\zeta}}

\newcommand{\Ex}{\mathbb{E}}

\ifdefined\macrosSB \else

\fi

\newcommand{\ddp}[2]{\frac{\partial #1}{\partial #2}}

\newcommand{\phib}{\bar\phi}

\ifdefined\macrosH
  \usepackage{xr-hyper}
  \usepackage{hyperref}
  \hypersetup{hypertexnames=false}
  \hypersetup{colorlinks,citecolor=blue,linkcolor=blue}  

\else\ifdefined\macrosHarxiv
  \usepackage{xr-hyper}
  \usepackage{hyperref}
  \hypersetup{hypertexnames=false, hidelinks}

\else
  
  \usepackage{xr}
\fi\fi

\usepackage[textwidth=500pt,textheight=660pt,centering]{geometry} 
\usepackage[dvips]{graphicx}
\usepackage{enumitem}

\usepackage[T1]{fontenc}

\hypersetup{
  urlcolor=black}

\newcommand{\Vdot}{\dot V}

\newcommand{\xib}{\bar{\xi}}
\newcommand{\Vpm}{Q}

\newcommand{\SD}{q}

\newcommand{\Xinew}{\mathrm{Z}}
\newcommand{\taunew}{+}

\begin{document}
\title{
Mean-field tricritical polymers
}

 \author{
   Roland Bauerschmidt\thanks{Department of Pure Mathematics and
     Mathematical Statistics, University of Cambridge,
     Wilberforce Road, Cambridge, CB3 0WB, UK.
   https://orcid.org/0000-0001-7453-2737.
     {\tt rb812@cam.ac.uk}}
    \and
   Gordon Slade\thanks{Department of Mathematics,
     University of British Columbia,
     Vancouver, BC, Canada V6T 1Z2.
     https://orcid.org/0000-0001-9389-9497.
     {\tt slade@math.ubc.ca}
     }}

\date{\vspace{-5ex}} 

\maketitle

\begin{abstract}
We provide an introductory
account of a tricritical phase diagram, in the setting of a mean-field random walk
model of a polymer density transition, and clarify the nature of the density
transition in this context.
We consider a continuous-time random walk model on the complete graph, in the limit as the number of vertices $N$ in the graph grows to infinity.
The walk has a repulsive self-interaction, as well as
a competing attractive self-interaction whose strength is controlled by a
parameter $g$.  A chemical potential $\nu$ controls the walk length.  We determine
the phase diagram in the $(g,\nu)$ plane, as a model of a density transition for a
single linear polymer chain.
A dilute phase (walk of bounded length)
is separated from a dense phase (walk of length of order $N$) by a phase boundary
curve.  The phase boundary is divided into two parts, corresponding to
first-order and second-order
phase transitions, with the division occurring at a tricritical point.
The proof uses a supersymmetric representation for the random walk model, followed
by a single block-spin renormalisation group step to reduce the problem to a
1-dimensional integral, followed by application of the
Laplace method for an integral with a large parameter.
\end{abstract}

%

\section{The model and results}

\subsection{Introduction}

Models of critical phenomena such as the Ising model and percolation
continue to be of central interest in the probability literature.
In such models, a single parameter (temperature for the
Ising model or occupation density for percolation) is tuned to a critical
value in order to observe universal critical behaviour.
In tricritical models, it is instead necessary to tune two parameters
simultaneously to observe tricritical behaviour.
Despite their importance for physical applications, tricritical phenomena
have received much less attention in the mathematical literature than
critical phenomena.  Our purpose in this paper is to provide an introductory
account of a tricritical phase diagram, in the setting of a mean-field random walk
model of a polymer density transition, and to clarify the nature of the density
transition in this context.

The self-avoiding walk is a starting point for the mathematical modelling of
the chemical physics of a single linear polymer chain in a solvent \cite{Genn79}.
The theory of the self-avoiding walk has primarily been developed
in the setting of an infinite lattice, often $\Z^d$.  So far, this theory has failed to
provide theorems capturing the critical behaviour in dimensions $d=2,3$,
such as a precise description of the typical end-to-end distance, and such problems
are rightly considered to be both highly important and notoriously difficult.
On $\Z^d$,
basic quantities such as the susceptibility---the generating function $\sum_{n=0}^\infty
c_n z^n$ for the number
of $n$-step self-avoiding walks started from the origin---can be used to model a polymer
chain in the \emph{dilute phase}.
The susceptibility is undefined when $|z|$ exceeds
the reciprocal of the connective constant $\mu=\lim_{n\to\infty}c_n^{1/n}$.
It is however large values of $z$ that
are required to model the \emph{dense phase}, as in \cite{BGJ05,Madr95a,DKY14}, and some
finite-volume approximation is needed for this.
Much remains to be learned about the phase transition from the dilute to the dense phase,
including its tricritical nature.

We study a mean-field model based on a continuous-time random walk on the complete graph
on $N$ vertices, in the limit $N\to\infty$.
The walk has a repulsive self-interaction which models the excluded-volume effect
of a linear polymer, as well as
a competing attractive self-interaction which models the tendency of the polymer to
avoid contact with the solvent.
The strength of the self-attraction is controlled by a
parameter $g$, with attraction increasing as $g$ becomes more negative.
A chemical potential $\nu$ controls the walk length.
We investigate
the phase diagram in the $(g,\nu)$ plane $\R^2$ (positive and negative values),
as a model of a density transition for a
single linear polymer chain.

\begin{figure}[ht]
\begin{center}
\input{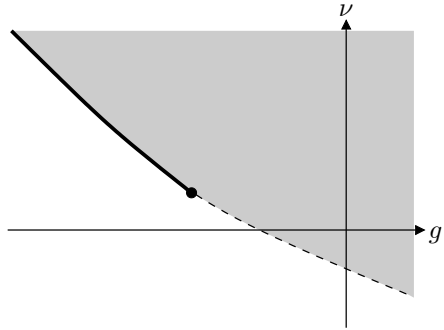}
\end{center}
\caption{
Typical tricritical phase diagram.
The second-order curve (dashed line) and the first-order curve (solid line) meet
at the tricritical point.  The shaded region is the dilute phase (bounded susceptibility)
and the unshaded region is the dense phase.}
\label{fig:generic}
\end{figure}

In the physics literature, the nature of the phase diagram is well understood.
The dilute and dense phases are separated by a phase boundary curve $\nu=\nu_c(g)$
as in Figure~\ref{fig:generic}.
The phase boundary itself is divided into two parts:  a second-order part for $g>g_c$
across which the average polymer
density varies continuously, and a first-order part for $g<g_c$
across which the density has a jump
discontinuity.  The two pieces of the phase boundary are separated by the tricritical point
$(g_c,\nu_c(g_c))$,
known as the \emph{theta point}.  Tricritical behaviour differs from critical behaviour
in the number of parameters that must be tuned.  For critical behaviour, an experimentalist
needs to tune a single variable to its critical value (given $g$, tune to $\nu_c(g)$).
For tricritical behaviour, two variables must be tuned (tune to $(g_c,\nu_c(g_c))$).
A mathematically rigorous theory of the mean-field tricritical polymer density
transition has been lacking, and our purpose here is to provide such a theory.
Our analysis could be extended to study the tricritical behaviour of $n$-component
spins or higher-order multi-critical points.
Surprisingly, the mean-field theory of the density transition for
the strictly self-avoiding walk has only very recently been developed \cite{Slad19,DGGNZ19}.

The upper critical dimension for the tricritical behaviour is predicted to be $d=3$,
and mean-field tricritical behaviour is predicted for the model on $\Z^d$ in
dimensions $d >3$.
On the other hand, for the critical behaviour associated with the second-order part of
the phase boundary, the upper critical dimension is instead $d=4$.

Nonrigorous methods were used in the physics literature to
study the density transition in dimensions 2 and 3, and in particular its tricritical behaviour,
in the 1980s \cite{DS87,Dupl86a,Dupl87,DS87a}.
In recent work with Lohmann, we applied a rigorous renormalisation group method
to study the
3-dimensional tricritical point \cite{BLS19}, and proved that the tricritical two-point
function has Gaussian $|x|^{-1}$ decay for the model on $\Z^3$.
In \cite{GI95}, the transition across
the second-order phase boundary was studied on a 4-dimensional hierarchical lattice,
where a logarithmic correction to the mean-field behaviour of the density was proved.
All of these references make use of an interpretation of the polymer model as the
$n=0$ version of an $n$-component spin model.  We also implement this strategy,
using an exact representation of the random walk model based on supersymmetry.
After a transformation which can be regarded as a single block-spin renormalisation
group step, this representation takes on a form which permits application of the Laplace method
for integrals involving a large parameter.

In the mathematical literature, it has been more common to model the polymer collapse
transition in terms of the \emph{interacting self-avoiding walk} in which
a walk with a self-repulsion receives an energetic reward for nearest-neighbour contacts.
A review of the literature on this model can be found in \cite[Chapter~6]{Holl09};
more recent papers include \cite{BSW-saw-sa,HH19,PT18}.
In our mean-field model set on the complete graph, there is no geometry,
and the notion of collapse (a highly localised walk) is not meaningful.
We therefore concentrate on the density transition and its
tricritical behaviour.

\subsection{The model}

\subsubsection{Definitions}

Let $\Lambda$ be a finite set with $N$ vertices; ultimately we are interested
in the limit $N \to \infty$.
Let $X=(X(t))_{t\in[0,\infty)}$ be the continuous-time simple random walk on the complete graph with
vertex set $\Lambda$.  This is the walk
with generator $\Delta$ defined, for $f: \Lambda \to \R$, by
\begin{equation}
\lbeq{Deltadef}
    (\Delta f)_x = \frac 1N \sum_{y\in\Lambda}(f_y-f_x)
    \qquad (x \in \Lambda).
\end{equation}
Equivalently, when the walk is at $x\in\Lambda$, it steps to a uniformly chosen
vertex in $\Lambda\setminus \{x\}$ after an exponentially distributed
holding time with rate $1-\frac{1}{N}$.  The steps and holding times are all
independent.
We denote expectation for $X$ with initial point $X(0)=x$ by $E_x$.

The \emph{local time} of $X$ at $x$ up to time $T$ is the random variable
\begin{equation}
    L_{T,x} = \int_0^T \1_{X(t)=x}\, dt
    \qquad (x\in\Lambda),
\end{equation}
which measures the amount of time spent by the walk at $x$ up to time $T$.
Let $L_T$ denote the vector of all local times.
Given a function $p:[0,\infty) \to [0,\infty)$ with $p(0)=1$,
we write $p_N(L_T)=\prod_{x\in \Lambda}p(L_{T,x})$.
Let $x,y\in\Lambda$.  Assuming the integrals exist, the \emph{two-point function} is
\begin{equation}
\lbeq{2ptfcndef}
    G_{xy}
    =
    \int_0^\infty E_x( p_N(L_T) \1_{X(T)=y}) \, dT
    ,
\end{equation}
and the \emph{susceptibility} is
\begin{equation}
    \chi = \sum_{y\in \Lambda} G_{xy}
    =
    \int_0^\infty E_x( p_N(L_T) ) \, dT
    .
\end{equation}
The right-hand side is independent of $x \in\Lambda$.

We define the random variable $L$, the \emph{length} of $X$, by its probability
density function
\begin{equation}
    f_L ( T) =
    \frac{1}{\chi}
    E_x\left(   p_N(L_T) \right)
    \qquad (T \ge 0)
    ,
\end{equation}
which is also independent of $x \in\Lambda$.
The expected value of the length is
\begin{equation}
\lbeq{EL0}
    \Ex L
    =
    \frac{1}{\chi}
    \int_0^\infty T E_x(p_N(L_T)) \, dT
    .
\end{equation}
The expected length can be written more compactly using a
dot to represent differentiation with respect to $\epsilon$ at $\epsilon=0$,
when $p(s)$ is replaced by $p(s)e^{-\epsilon s}$.
With this notation, since $T=\sum_{x\in\Lambda}L_{T,x}$,
\begin{equation}
\lbeq{ELdot}
    \Ex L
    =
    - \frac{1}{\chi} \dot{\chi} .
\end{equation}
Assuming the limit exists, the \emph{density} of the walk is defined by
\begin{equation}
    \rho  = \lim_{N \to\infty} \frac 1N \Ex L.
\end{equation}

\subsubsection{Example}

Although our results will be presented more generally,
we are motivated by the example
\begin{equation}
\lbeq{p-example}
    p(t) = e^{-t^3-gt^2-\nu t} \qquad (t \ge 0),
\end{equation}
where $g,\nu\in \R$ (we have set the coefficient of $t^3$ to equal $1$, its specific
value is unimportant).
For $p$ defined by \refeq{p-example}, the two-point function becomes
\begin{equation}
    G_{xy}(g,\nu) =  \int_0^\infty E_x(e^{-\sum_{x\in\Lambda}(L_{T,x}^3+gL_{t,x}^2)}
    \1_{X(T)=y})
    e^{-\nu T}  \, dT.
\end{equation}
The above integral is finite for all $g,\nu\in \R$, since by H\"older's inequality
$T = \sum_{x\in \Lambda} L_{T,x}
\le (\sum_{x\in\Lambda}L_{T,x}^3)^{1/3}|\Lambda|^{2/3}$, and also
$\sum_{x\in\Lambda}L_{T,x}^2 \le (\sup_x L_{T,x})\sum_x L_{T,x} \le T^2$, so
\begin{equation}
    G_{xy}(g,\nu) \le \int_0^\infty e^{-T^3|\Lambda|^{-2} + |g| T^2 +|\nu|T}\, dt < \infty .
\end{equation}

By definition,
\begin{align}
    \sum_{x \in \Lambda}L_{T,x}^2
    &= \int_0^T \int_0^T \1_{X(s)=X(t)} \, ds \, dt,
    \\
    \sum_{x \in \Lambda}L_{T,x}^3
    &= \int_0^T \int_0^T \int_0^T \1_{X(s)=X(t)=X(u)} \, ds \, dt  \, du.
\end{align}
Our interest lies in the case $g<0$.  In this case, walks $X$ for which the local time
has large $\ell^3$-norm are penalised by the factor $e^{-\sum_{x\in\Lambda}L_{T,x}^3}$
(three-body repulsion),
whereas those with large $\ell^2$-norm are rewarded by the factor
$e^{+\sum_{x\in\Lambda}|g|L_{T,x}^2}$ (two-body attraction).
This is a model of a linear polymer in a solvent.
The parameter $\nu$ is a chemical potential which controls the length
of the polymer.  The three-body repulsion models the excluded volume effect, and
the two-body attraction models the effect of temperature or solvent quality.
The competition between attraction and repulsion, together with the variable length
mediated by the chemical potential, leads to a rich phase diagram.

\subsubsection{Effective potential}

The mean-field Ising model, known as the Curie--Weiss model, can be analysed
in terms of the effective
potential $V_{{\rm Ising}}(\varphi) = \frac{\beta}{2}\varphi^2 - \log\cosh(\beta \varphi)$.
In \cite[Section~1.4]{BBS-brief},
this effective potential was derived as the result of a single block-spin
renormalisation group step.  Our approach is based on this idea.

For the mean-field polymer model with interaction $p$,
we define the \emph{effective potential} $V: [0,\infty) \to \R$  by
\begin{align}
    V(t) &= t - \log(1+v(t)),
\qquad
    v(t)  =
    \int_0^\infty p(s)e^{-s} \sqrt{\frac{t}{s}} I_1(2\sqrt{st}) \, ds,
\lbeq{vintegral}
\end{align}
with $I_1$ the modified Bessel function of the first kind.
We show in Proposition~\ref{prop:Vderivs0} that the integral $v(t)$ is finite
if $p$ is integrable.
By definition, $V(0)=0$.

The variable $t$ corresponds to $\frac12 \varphi^2$ for the Ising effective potential.
It is common in tricritical theory to encounter a triple-well potential;
a double well for $V(t)$ in our setting corresponds to a triple well as a function
of $\varphi$.

The effective potential occurs in integral representations of the two-point function,
the susceptibility, and the expected length.  In contrast to the analysis of
the mean-field Ising model in \cite[Section~1.4]{BBS-brief}, the integral representations
involve the notions of fermions and supersymmetry as presented in
\cite[Chapter~11]{BBS-brief}.  Nevertheless, the integral representation reduces
to a 1-dimensional Lebesgue integral.  For example,
as noted below \refeq{G01bis},
the two-point function at distinct
points labelled $0,1$ has the integral representation
\begin{equation}
\lbeq{G01example}
    G_{01} =
    \int_0^\infty e^{-NV(t)} \Big(NV'(t) (1-V'(t))  + 2V''(t) \Big)(1-V'(t)) t \, dt.
\end{equation}
Similarly $G_{00}$, $\chi$, and $\Ex L$ are represented by integrals of the
form $\int_0^\infty e^{-NV(t)} K(t) \, dt$ for suitable kernels $K$.
The asymptotic behaviour of such integrals,
as $N \to \infty$, can be computed using the Laplace method.
This requires knowledge
of the minimum structure of the effective potential $V$.
The use of the minimum structure to predict the phase diagram is referred to as the
Landau theory (see, e.g., \cite[Section~7.6.4]{Arov18} where our variable $t$ corresponds
to $m^2$).

We assume throughout the paper that $p: [0,\infty)\to[0,\infty)$ is
such that $e^{\epsilon s} p(s)$ is a Schwartz
function for some $\epsilon >0$; this assumption is a convenience that
permits direct application
of the integral representation given in Theorem~\ref{thm:BFS-Dynkin}.
In particular, $p$ is integrable.
We also assume that $p(0)=1$.

The following elementary proposition collects some basic facts about the
effective potential.  Weaker assumptions on $p$ and a stronger analyticity conclusion
for $V$
are possible, but the proposition is sufficient for our needs as it is stated.
The proposition shows that $V$ is analytic on $(0,\infty)$ under our assumption that $p$ is integrable.  It also shows that the
derivatives of the effective potential at $t=0$ can be expressed in terms
of the moments of $p(s)e^{-s}$ defined by
\begin{equation}
\lbeq{pmoments}
    M_k = \int_0^\infty p(s)e^{-s} s^{k}   ds  \qquad (k=0,1,\ldots).
\end{equation}
Such derivatives appear in Definition~\ref{defn:phases} and
Theorems~\ref{thm:2ptfcn-mr}--\ref{thm:mr} below.

\begin{prop}
\label{prop:Vderivs0}
(i) If $\int_0^\infty p(s) \, ds<\infty$  then $V$ is well-defined and analytic in
$t\in (0,\infty)$.
Moreover, if $p(s) \leq O(e^{-\epsilon s})$ for some $\epsilon>0$ then there
exists $\delta>0$ such that
$V(t) \geq \delta t + O(1)$ as $t\to\infty$.
\\
(ii)
Derivatives of $V$ at $t=0$ (with the dot notation as indicated above \refeq{ELdot})
are given by
\begin{gather} \label{e:Vderivs0}
    V'(0)  = 1 - M_0 ,  \quad V''(0) = M_0^2 -M_1,
    \quad
    V'''(0) = -\frac 12 M_2 + 3M_1M_0 -2 M_0^{3},
    \\
    \Vdot(0)  = 0, \quad
    \Vdot'(0) = M_1.
\end{gather}
\end{prop}

\begin{proof}
(i)
The modified Bessel function $I_1$ is
the entire function
$I_1(z) =  \sum_{k=0}^\infty \frac{1}{k!(k+1)!} ( \frac{z}{2} )^{2k+1}$.
Its asymptotic behaviour is $I_1(z) \sim \frac{z}{2}$
as $z \downarrow 0$ and $I_1(z) \sim \frac{1}{\sqrt{2\pi z}}e^z$ as $z\to\infty$.
Thus the integral $v(t)$ converges when $p$ is integrable, and $V(t)$ is well-defined.

To prove that $V$ is analytic on $(0,\infty)$,
since $t \mapsto \sqrt{t}$ is analytic, it suffices to prove that
the function $w(z)=v(z^2)$ is entire.  By definition, for $z \in \C$,
\begin{equation}
    w(z) = z \int_0^\infty p(s) e^{-s}  s^{-1/2} I_1(2s^{1/2} z) \, ds.
\end{equation}
The modified Bessel functions obey
  $|I_n(z)| \le I_n(|z|) \sim \frac{1}{\sqrt{2\pi |z|}}e^{|z|}$ as $|z|\to\infty$
and the same asymptotics hold for their derivatives.
This permits differentiation under the integral and guarantees existence of the derivative $w'(z)$.

For the lower bound on $V$, we apply
the assumption $p(s) \leq O(e^{-\epsilon s})$, the bounds $I_1(z) \leq O(e^{z})$
for $z \geq 1$ and $I_1(z)\leq O(z)$ for $z\leq 1$,
and the inequality $2\sqrt{st} \leq
(1+\epsilon/2)s + (1-\delta)t$ with $1-\delta=(1+\epsilon/2)^{-1}$.
Together, these lead to
\begin{equation}
  v(t)
  \leq
  C +
  C \int_1^\infty e^{-s-\epsilon s}  e^{2\sqrt{st}} \, ds
  \leq C e^{(1-\delta) t} \int_0^\infty e^{-\epsilon s/2} \, ds
  \leq O(e^{(1-\delta) t}),
\end{equation}
and hence $V(t) = t - \log(1+v(t)) \geq \delta t+O(1)$.

\smallskip\noindent
(ii) The Taylor expansion of $v(t)$ at $t=0$ is
\begin{align} \label{e:v-expansion}
    v(t) & = \int_0^\infty p(s)e^{-s}
    \left( \sum_{k=0}^\infty \frac{1}{k!(k+1)!} t^{k+1}s^{k} \right) ds
    =
    \sum_{k=0}^\infty \frac{1}{k!(k+1)!} M_k t^{k+1}.
\end{align}
In particular,
\begin{equation}
\lbeq{vderivs}
    v(0)=0, \qquad v'(0) = M_0, \qquad v''(0) = M_1, \qquad
    v^{(k)}(0) = \frac{M_{k-1}}{(k-1)!} \quad (k \ge 1).
\end{equation}
Computation gives
\begin{equation}
    V'(t) = 1 - \frac{v'(t)}{1+v(t)},
    \qquad
    V''(t) = - \frac{v''(t)}{1+v(t)} +  \left( \frac{v'(t)}{1+v(t)} \right)^2,
\end{equation}
and the third derivative can be computed similarly.
This leads to the statements for the derivatives of $V$ with respect to $t$.

Finally, $\Vdot(0)=0$ since $V(0)=0$ holds also when $p(s)$ is replaced by
$p(s)e^{-\epsilon s}$, and
$\Vdot'(0)=M_1$ follows from $V'(0)=1-M_0$ and $\dot{M}_0=-M_1$.
\end{proof}

\subsection{General results}
\label{sec:results}

In the following definition, we have in mind the situation where
the effective potential $V$ is defined by a function $p$ which is
parametrised by two real parameters $(g,\nu)$
as in \refeq{p-example}.
Different choices of parameters can correspond to different cases
in the definition.
For the specific example
of \refeq{p-example}, plots of the
phase diagram and effective potential are given in Figures~\ref{fig:nuc}--\ref{fig:Veg}.
However, our results and their proofs depend only on the qualitative features of the effective
potential listed in the definition.

We say that $V$ \emph{has a unique global minimum} $V(t_0)$
if: (i) $V(t)>V(t_0)$ for all
$t\neq t_0$, and (ii) $\inf_{t\in [0,\infty) : |t-t_0|\ge \epsilon}(V(t)-V(t_0))>0$
for all $\epsilon >0$.
We say that $V$ \emph{has global minima} $V(t_0)=V(t_1)$ with $t_0\neq t_1$ if:
(i)  $V(t)>V(t_0)=V(t_1)$ for all
$t\neq t_0,t_1$, and (ii)
$\inf_{t\in [0,\infty) : |t-t_0|\ge \epsilon,\, |t-t_1|\ge \epsilon}(V(t)-V(t_0))>0$
for all $\epsilon >0$.

\begin{defn} \label{defn:phases}
We define two phases, two phase boundaries, and the tricritical point,
in terms of the effective potential $V$ as follows:
\begin{align*}
    \text{dilute phase:} &\quad V'(0)>0,  \;
    \text{unique global minimum $V(0)=0$}.
    \nnb
    \text{second-order curve:} &\quad V'(0)=0, \; V''(0)> 0,
     \; \text{unique global minimum $V(0)=0$}.
    \nnb \text{tricritical point:} & \quad V'(0)=V''(0)=0, \;
    V'''(0)>0,   \; \text{unique global minimum $V(0)=0$}.
    \nnb
    \text{first-order curve:} &\quad
    \text{$V'(0)>0$, global minima $V(0)=V(t_0)=0$ with $t_0>0$,}\;
    V''(t_0)>0  .
    \nnb
    \text{dense phase:} &\quad
    \text{unique global minimum}\;
    V(t_0)<0 \;  \text{with}
    \;  t_0>0,\;  V''(t_0)>0.
\end{align*}
\end{defn}

In principle, there are further possibilities such as $n^{\rm th}$-order critical points.
  These do not occur in our example \refeq{p-example}, so we do not consider them,
  but they could be handled in an analogous way.

The following two theorems give the asymptotic behaviour of the two-point
function, the susceptibility, and the expected length, in the different regions
of the phase diagram.  The result for the susceptibility is a consequence
of the result for the two-point function, together with the identity
$\chi = G_{00}+(N-1)G_{01}$.
As usual, the Gamma function is $\Gamma(x) =
\int_0^\infty t^{x-1} e^{-t} dt$ for $x>0$.
The notation $f(N) \sim g(N)$ means $\lim_{N\to\infty}f(N)/g(N)=1$.

\begin{theorem}
\label{thm:2ptfcn-mr}
The two-point function has the asymptotic behaviour:
\begin{align}
\lbeq{G00asy-mr}
    G_{00} & \sim
    \begin{cases}
    1-V'(0) & \quad\text{(dilute phase and first-order curve)}
    \\
    1 & \quad\text{(second-order curve)}
    \\
    1 & \quad\text{(tricritical point)}
    \\
    e^{N|V(t_0)|}\frac{1}{N^{1/2}}
    \frac{\sqrt{2\pi}}{V''(t_0)^{1/2}}(1-V''(t_0))
    & \quad
    \text{(dense phase)},
    \end{cases}
\\
\lbeq{G01asy-mr}
    G_{01} & \sim
    \begin{cases}
    \frac{(1-V'(0))^2}{V'(0)N}
    & \qquad\qquad\qquad\text{(dilute phase)}
    \\
    \frac{1}{(\frac {1}{2!} V''(0)N)^{1/2}} \Gamma(3/2) & \qquad\qquad\qquad\text{(second-order curve)}
    \\
    \frac{1}{(\frac {1}{3!} V'''(0)N)^{1/3}} \Gamma(4/3) & \qquad\qquad\qquad\text{(tricritical point)}
    \\
    e^{N|V(t_0)|}\frac{1}{N^{1/2}} \frac{\sqrt{2\pi}}{V''(t_0)^{1/2}}
    & \qquad\qquad\qquad
    \text{(dense phase and first-order curve)}.
    \end{cases}
\end{align}
\end{theorem}

The statement of the next theorem uses the notation $\Vdot(t_0)$ and $\Vdot'(0)$.
The dot notation is as discussed above \refeq{ELdot}.  Explicitly,
\begin{equation}
\lbeq{Vdotdef}
    \Vdot (t) = -\frac{\dot{v}(t)}{1+v(t)},
    \qquad
    \dot{v}(t) = -\int_0^\infty p(s)e^{-s} \sqrt{st} I_1(2\sqrt{st}) \, ds,
\end{equation}
and, as usual, a prime denotes differentiation with respect to $t$.

\begin{theorem}
\label{thm:mr}
The susceptibility $\chi$ and expected length $\Ex L$
have the asymptotic behaviour:
\begin{align}
\lbeq{chiasy-mr}
    \chi & \sim
    \begin{cases}
    \frac{1-V'(0)}{V'(0)}  & \text{(dilute phase)}
    \\
    N^{1/2}
     \frac{\Gamma(3/2)}{(\frac{1}{2!}V''(0))^{1/2}} & \text{(second-order curve)}
    \\
    N^{2/3}\frac{\Gamma(4/3)}{(\frac {1}{3!} V'''(0))^{1/3}}
    & \text{(tricritical point)}
    \\
    e^{N |V(t_0)|}
    N^{1/2}\frac{\sqrt{2\pi}}{V''(t_0)^{1/2}}   & \text{(dense phase and first-order curve),}
    \end{cases}
\\
\lbeq{ExLasy-mr}
    \Ex L & \sim
    \begin{cases}
    \frac{\Vdot'(0)}{V'(0)(1-V'(0))}  & \text{(dilute phase)}
    \\
    N^{1/2}
    \frac{1}{\Gamma(1/2)}   \frac{\Vdot'(0)}{(\frac{1}{2!}V''(0))^{1/2}}
     & \text{(second-order curve)}
    \\
    N^{2/3}  \frac{\Gamma(2/3)}{\Gamma(1/3)}
         \frac{1}{(\frac{1}{3!}V'''(0))^{1/3}}
       & \text{(tricritical point)}
    \\
     N\Vdot(t_0)   & \text{(dense phase and first-order curve).}
    \end{cases}
\end{align}
\end{theorem}

By Theorem~\ref{thm:2ptfcn-mr},
the two-point function remains bounded in the dilute phase, on the
first- and second-order curves,
and at the tricritical point.  Also, $G_{00}$ is asymptotically constant
in the dilute phase, on the second-order curve, and at the tricritical point, whereas
$G_{01}$ decays at different rates in the different regions.
In the dense phase, both $G_{00}$ and $G_{01}$ grow exponentially in $N$.
Proposition~\ref{prop:Vderivs0}(ii) shows that in all cases $1-V'(0)>0$,
as is implied in particular in the dilute phase by the first asymptotic formula for $G_{00}$.
The formula for $G_{00}$ in the dense phase implies that $V''(t_0)<1$; we do not
have an independent general proof of that (though if $t_0$ is smooth in $(g,\nu)$
then it is true in the vicinity
of the tricritical point where $t_0=0$ and $V''(0)=0$).

Theorem~\ref{thm:mr} indicates that the susceptibility $\chi$ and expected length $\Ex L$
each have
finite infinite-volume limits in the dilute phase.
In the dense phase, $\chi$ grows exponentially with $N$.
In the dense phase and on the first-order curve,
$\Ex L$ is asymptotically linear in $N$.
On the second-order curve, $\chi$ and $\Ex L$
are each of order $N^{1/2}$ (as in \cite{Slad19,DGGNZ19} for the self-avoiding walk
on the complete graph), whereas each is of order
$N^{2/3}$ at the tricritical point.
The density
$\rho  = \lim_{N \to\infty} N^{-1} \Ex L$
is zero except on the first-order curve and in the dense phase, where it is equal
to $\Vdot(t_0)>0$.

The proofs of Theorems~\ref{thm:2ptfcn-mr}--\ref{thm:mr} are given in two steps.
In Section~\ref{sec:susy}, integral representations based on supersymmetry are derived;
these involve the effective potential.
In Section~\ref{sec:pf-mr}, the Laplace method is used to evaluate the asymptotic
behaviour of the integrals.

\subsection{Phase diagram for the example}
\label{sec:intro-example}

For further interpretation of the phase diagram, we restrict attention in this
section to the particular example
\begin{equation}
\lbeq{pexample}
    p(t) = e^{-t^3-gt^2-\nu t},
\end{equation}
for which we carry out numerical calculations to determine the structure of the effective potential.
The numerical input we need is collected in
Section~\ref{sec:numericalinput}, and we mention some of it here.

\begin{figure}[ht]
\centering{
\includegraphics[scale=1.0]{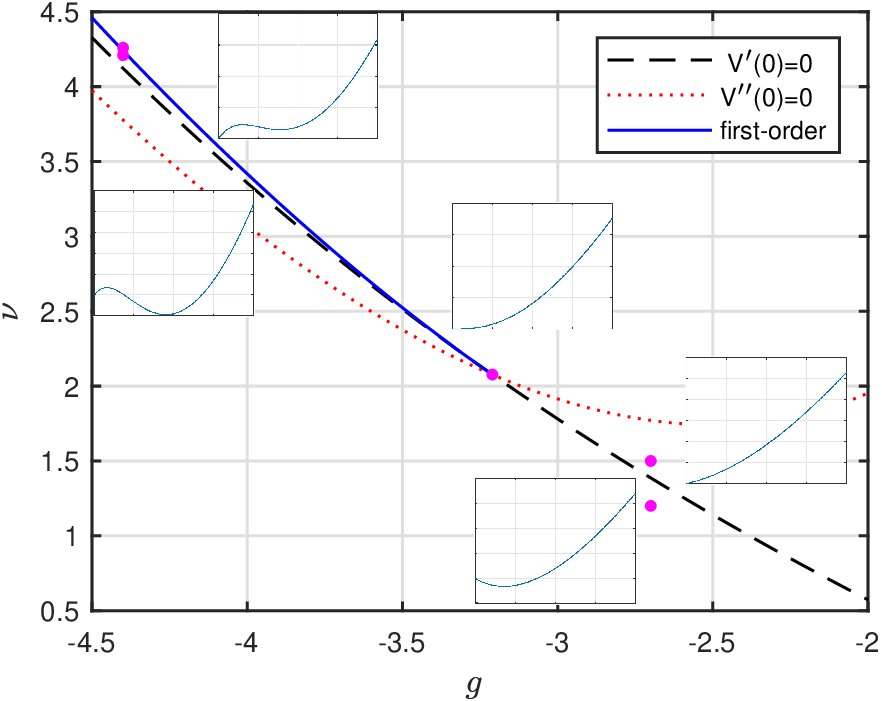}
\caption{Phase diagram for $p(t) = e^{-t^3-gt^2-\nu t}$.  The five marked points
with their effective potentials are those shown in Figure~\ref{fig:Veg}.}
\label{fig:nuc}
}
\end{figure}

Two curves which provide bearings
in the $(g,\nu)$ plane are determined by the equations
$V'(0)=0$ (i.e., $M_0=1$) and $V''(0)=0$ (i.e., $M_1=M_0^2$).
The curves, which are plotted in Figure~\ref{fig:nuc},
intersect at the tricritical point (i.e., $M_0=M_1=1$), which is
\begin{equation}
  g_c = -3.2103..., \quad \nu_c = 2.0772... \, .
\end{equation}
At the tricritical point, numerical integration gives $M_2 = 1.4478...$ and
$V'''(0) = 0.2762...>0$.
The first-order curve is the
blue (solid) curve in Figure~\ref{fig:nuc}.
The second-order curve is the portion of the black (dashed) curve below the tricritical point.
The dilute phase lies above the first- and second-order curves, and the dense phase
comprises the other side of those curves.
The \emph{phase boundary} is the union of the first- and second-order curves together
with the tricritical point; we regard this curve as a function $\nu_c(g)$
parametrised by $g$.

By Theorem~\ref{thm:mr}, there is a transition as the phase boundary
$g \mapsto (g,\nu_c(g))$ is traversed
in the direction of decreasing $g$:
\begin{itemize}
\item
On the second-order curve, $\chi$ and $\Ex L$ are of order $N^{1/2}$ and $\rho=0$.
\item
At the tricritical point, $\chi$ and $\Ex L$ are of order $N^{2/3}$ and $\rho=0$.
\item
On the first-order curve, $\chi$
is of order $N^{1/2}$, $\Ex L$ is of order $N$,
and $\rho = \Vdot(t_0)>0$.
\end{itemize}
This  is a density transition, from zero to
positive density.
Note that, by definition, $\Vdot = \ddp{V}{\nu}$ when $p$ is given by \refeq{pexample}.

\begin{figure}[ht]
\centering{
\includegraphics[scale=0.8]{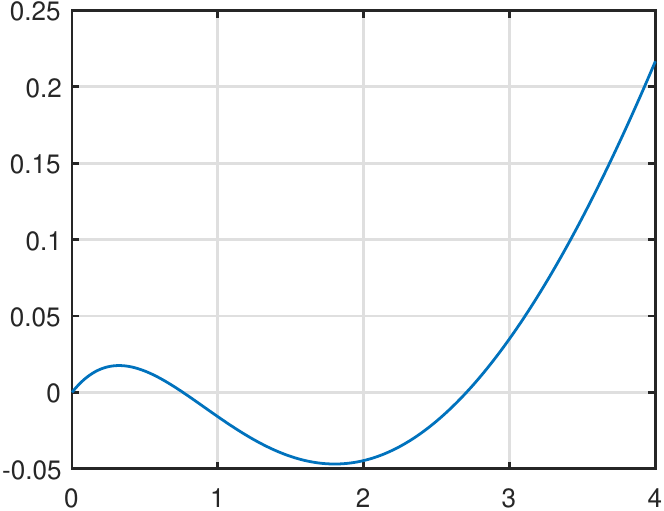}
\includegraphics[scale=0.8]{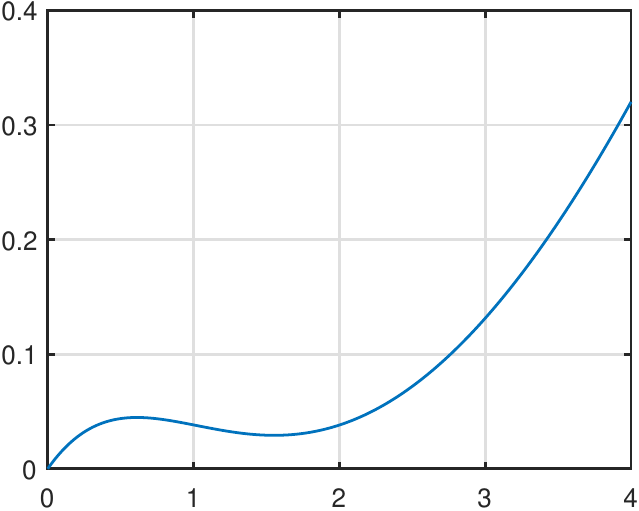}

\scriptsize{
dense phase near first-order curve $(-4.4,4.21)$
\hspace{15mm} dilute phase near first-order curve $(-4.4,4.26)$}

\includegraphics[scale=0.8]{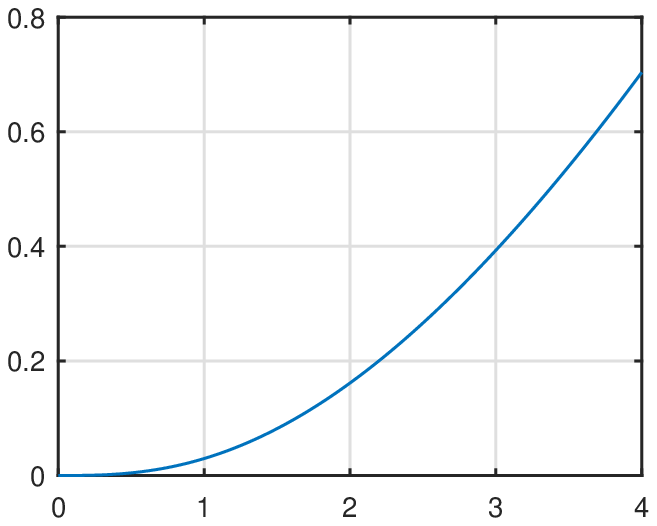}

\scriptsize{
tricritical point}

\includegraphics[scale=0.8]{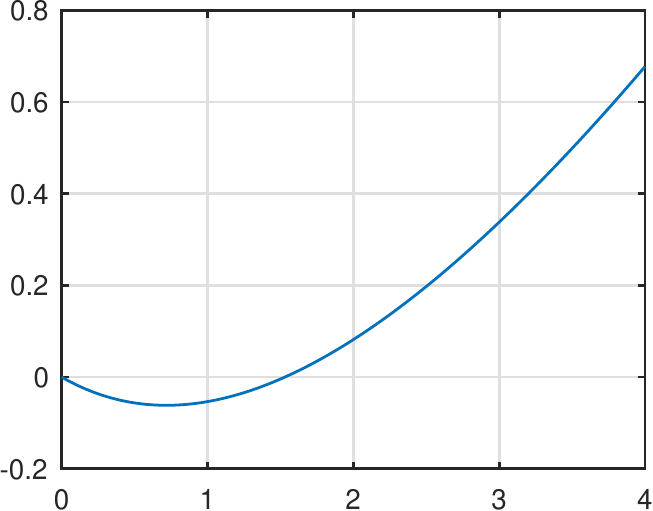}
\includegraphics[scale=0.8]{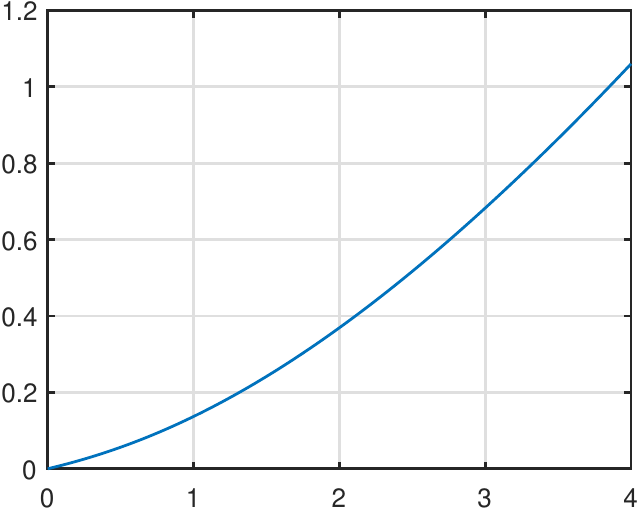}

\scriptsize{
dense phase near second-order curve $(-2.7,1.2)$
\hspace{15mm} dilute phase near second-order curve $(-2.7,1.5)$}

\caption{Effective potential $V$ vs $t$ for several values of $(g,\nu)$.}
\label{fig:Veg}}
\end{figure}

\smallskip\noindent\emph{First-order curve.}
The density $\rho$
is discontinuous when crossing the first-order curve,
since its value on the first-order curve is $\Vdot(t_0)>0$
whereas its value in the dilute phase is zero. However, the density
is continuous at
the first-order curve for the one-sided approach from the dense phase.
This can be understood from the behaviour of the effective potential:  as the first-order
curve is approached from the dense phase, $t_0$ remains bounded away from zero
and $\Vdot(t_0)$ does not vanish (upper two images
in Figure~\ref{fig:Veg}).
The density discontinuity on the first-order curve
 is in contrast to the continous behaviour on the second-order curve.
As the second-order curve (or tricritical point) is approached from the dense phase,
$t_0$ decreases continuously to zero and $\Vdot(t_0) \downarrow 0$ (lower two images
in Figure~\ref{fig:Veg}).

In the limit $N\to\infty$,
the susceptibility has finite limit $\frac{1-V'(0)}{V'(0)}$ as the first-order curve is approached from
the dilute phase, whereas it is divergent on the first-order curve.
This is typical of a first-order transition.

\smallskip\noindent\emph{Second-order curve and tricritical point.}
The detailed asymptotic behaviour of the divergence of the susceptibility and
the vanishing of the density, at the second-order
curve and tricritical point, are as
described in the following theorem.
The theorem also describes the phase boundary at the tricritical point.
Its proof is given in Section~\ref{sec:phase-diagram-pfs}.

\medskip
Theorem~\ref{thm:exponents} relies on numerical analysis of the effective potential
for $p$ given by \refeq{pexample}, as discussed above.
The precise conclusions from this numerical analysis are stated in Section~\ref{sec:numericalinput}.
We emphasise that the effective potential is a function of a single real variable,
and thus we believe that with effort this numerical input could
be replaced by rigorous analysis (perhaps with computer assistance), but we do not pursue this.

In the theorem, we consider a line segment
\begin{equation} \label{e:segment-bis}
    (g(s),\nu(s)) = (g(0)+sm_1,\nu(0)+sm_2)
    \qquad (s\in [0,1])
\end{equation}
that approaches a base point $(g(0),\nu(0))$ as $s\downarrow 0$.
We write ${\bf m} = (m_1,m_2)$ for its direction.
The base point may be either the tricritical point or a point on the second-order curve.

\begin{theorem}
\label{thm:exponents}
Let $p$ be given by \refeq{pexample}.

\smallskip\noindent
(i) The phase boundary $\nu_c(g)$ is differentiable with respect to $g$ at the
tricritical point, with slope $-M_2$.
However, its left and right second derivatives differ at the tricritical point.

\smallskip\noindent
(ii)
The vector ${\bf n} = (M_2,M_1)$ is normal along the second-order curve.
Along a line segment \eqref{e:segment-bis} approaching a point
on the second-order curve, or the tricritical point,
from the dilute phase with direction
satisfying ${\bf m} \cdot {\bf n} \neq 0$
(nontangential at the given point), the infinite-volume susceptibility
$\chi = \frac{1-V'(0)}{V'(0)}$
diverges as
\begin{equation}
\lbeq{chiasy-thm}
\chi = \frac{1-V'(0)}{V'(0)}
\sim \frac{1}{|{\bf m} \cdot {\bf n}| s}.
\end{equation}
(iii)
Along a line sement \eqref{e:segment-bis} approaching a point
on the second-order curve from the dense phase
with direction
satisfying ${\bf m} \cdot {\bf n} \neq 0$ (nontangential at the given point),
the density $\rho$
vanishes as
\begin{equation}
\lbeq{rhoasy2-thm}
    \rho \sim
    \frac{M_1}{1-M_1}|{\bf m}\cdot {\bf n}|s
    \qquad \text{(here $1-M_1>0$).}
\end{equation}
There exists an arc of the second-order curve adjacent to the tricritical
point, such that under tangential approach to a point on that arc,
$\rho \sim Bs^2$ with $B>0$.
\\
There are positive constants $B_0,B_1,B_2,B_3$ such that
as the tricritical point is approached,
\begin{equation}
\lbeq{rhoasy3-thm}
    \rho \sim
    \begin{cases}
    B_0 {(|{\bf m}\cdot {\bf n}|} s)^{1/2} & \text{(nontangentially from dense phase)}
    \\
    B_1 s & \text{(tangentially from second-order side)}
    \\
    B_2 s & \text{(tangentially from first-order side)}
    \\
    B_3 s & \text{(along first-order curve).}
    \end{cases}
\end{equation}
(For the first-order curve, the parametrisation is $(g(s),\nu(s))=(g_c-s,\nu_c(g-s))$.)
\end{theorem}

It is possible in general that the  susceptibility could have different asymptotic behaviour
for the approaches to the second-order curve and the tricritical point,
but for the mean-field model there is no difference.
However, for the density there is a difference.

\section{Integral representation}
\label{sec:susy}

In this section, we prove integral representations for the two-point
function and expected length, in Propositions~\ref{prop:2ptfcn}--\ref{prop:EL}, via the
supersymmetric version of the BFS--Dynkin isomorphism theorem \cite[Corollary~11.3.7]{BBS-brief}.
These integral representations are in terms of the effective potential and provide the
basis for the proofs of Theorems~\ref{thm:2ptfcn-mr}--\ref{thm:mr}.
We begin with brief background concerning Grassmann integration.
Further background and history for the isomorphism theorem
can be found in \cite[Chapter~11]{BBS-brief}.

\subsection{Grassmann algebra and the integral representation}
\label{sec:integration}

\subsubsection{Grassman algebra}

We define a Grassmann algebra $\Ncal_1$ with two generators
$\psi,\psib$ (the bar is only notational and is not a complex conjugate)
to consist of linear combinations
\begin{equation}
  K=a_0 + a_1\psib + a_2 \psi + a_3\psib\psi,
\end{equation}
where each $a_i$ is a smooth function $a_i:\R^2 \to \R$ written
$(u,v) \mapsto a_i(u,v)$, and where multiplication of the
generators is anti-commutative, i.e.,
\begin{equation}
  \psib\psi = -\psi\psib, \qquad \psi\psi = 0, \qquad \psib\psib = 0.
\end{equation}
To make the notation more symmetric, we also combine $(u,v) \in \R^2$ into a complex variable $\phi$ by
\begin{equation}
  \phi = u+iv, \qquad \phib = u-iv.
\end{equation}
We call $(\phi,\phib)$ a \emph{bosonic variable},
$(\psi,\psib)$ a \emph{fermionic variable},
and $\Phi = (\phi,\phib,\psi,\psib)$ a \emph{supervariable}.
Elements of the Grassmann algebra $\Ncal_1$ are called \emph{forms}.
A form with $a_1=a_2=0$ is called \emph{even}.
An important even form is
\begin{equation}
\lbeq{Phi2}
  \Phi^2
  = \phi\phib+\psi\psib.
\end{equation}

The above discussion concerns a single boson pair and a single fermion pair.
We also have need of the Grassmann algebra $\Ncal_N$ with $2N$
anticommuting generators
$(\psi_x,\psib_x)_{x\in\Lambda}$,
now with coefficients which are smooth functions
from $\R^{2N}$ to $\R$.
The \emph{even subalgebra} consists of elements of $\Ncal_N$ which only involve terms
containing products of an even number of generators.
We refer to $(\phi_x,\phib_x)_{x\in\Lambda}$ and $(\psi_x,\psib_x)_{x\in\Lambda}$ as the
boson field and the fermion field, respectively.
The combination $\Phi =(\phi_x,\phib_x,\psi_x,\psib_x)_{x\in\Lambda}$
is called a \emph{superfield}, and we write
\begin{equation}
    \Phi^2 = (\Phi^2_x)_{x\in\Lambda} =
    (\phi_x\phib_x + \psi_x\psib_x)_{x\in\Lambda}.
\end{equation}
Two useful even forms in $\Ncal_N$ are
\begin{align}
  (\Phi,\Phi) &= \sum_{x\in\Lambda}\Phi_x^2
  = \sum_{x\in\Lambda} (\phi_x\phib_x + \psi_x\psib_x),
  \\
  (\Phi,-\Delta\Phi) &= \sum_{x\in\Lambda}
  \Big(
  \phi_x(-\Delta \phib)_x+\psi_x (-\Delta \psib)_x
  \Big),
\end{align}
where $-\Delta$ is still defined by \refeq{Deltadef} when applied to the
generators $\psi$ and $\psib$.

For $p \in \N$, consider a $C^\infty$ function $F : \R^{p} \to\R$.
Let $K=(K_j)_{j \le p}$ be a collection of even forms, and assume
that the degree-zero part $K_j^0$ of each $K_j$ (obtained by setting all fermionic
variables to zero)
is real.  We define a form denoted $F(K)$ by Taylor series about the degree-zero
part of $K$, i.e.,
\begin{equation}
\label{e:Fdef}
    F(K) = \sum_{\alpha} \frac{1}{\alpha !}
    F^{(\alpha)}(K^{0})
    (K - K^{0})^{\alpha}.
\end{equation}
Here $\alpha = (\alpha_j)_{j \le p}$ is a multi-index, with $\alpha ! =
\prod_{j=1}^{p}\alpha_j !$ and $(K - K^{0})^{\alpha}
=\prod_{j=1}^{p} (K_{j} - K_{j}^{0})^{\alpha_{j}}$.  The order of
the product is immaterial since each $K_j-K_j^0$ is even by
assumption.  Also, the summation terminates
after finitely many terms since each $K_j-K_j^0$ is nilpotent.

For example, for $\Phi^2 \in \Ncal_1$ given by \refeq{Phi2}, for
smooth $F:\R\to\R$, the previous definition with $p=1$ gives
\begin{equation}
  F(\Phi^2)
  = F(\phi\phib) + F'(\phi\phib)\psi\psib
  = F(\phi\phib) - F'(\phi\phib)\psib\psi
    .
\label{e:ftau}
\end{equation}

\subsubsection{Grassmann integration and the integral representation}

Given  a form $K \in \Ncal_N$, we write $K_{2N}$ for its coefficient of
$\psib_1\psi_1\cdots\psib_N\psi_N$.
This $K_{2N}$ is a function of $(u,v)$, i.e., a function on $\R^{2N}$.
For example, for the form $K=F(\Phi^2)$ of \eqref{e:ftau}, we have $N=1$ and
$K_{2}(u,v) = -F'(\phi\phib) = -F'(u^2+v^2)$.
In general, the \emph{superintegral} of $K$ is defined by
\begin{equation}
\lbeq{intdef}
\int_{\R^{2N}} D\Phi \, K
= \frac{1}{\pi^N}\int_{\R^{2N}} K_{2N}(u,v) \, du\, dv,
\end{equation}
assuming that $K_{2N}$ decays sufficiently rapidly that the Lebesgue integral on the
right-hand side exists.  The notation $D\Phi$ signifies that $K$ is a form
for the superfield
$\Phi = (\phi_x,\phib_x,\psi_x,\psib_x)_{x\in \Lambda}$.
This will be useful to distinguish superfields when more than one are in play.
The factor $\pi^{-N}$ in the definition simplifies the
conclusions of the next example and theorem.

\begin{example}
If $F:\R \to \R$ decays sufficiently rapidly then
\begin{equation}
\lbeq{intftau}
    \int_{\R^2} D\Phi \, F(\Phi^2) = F(0).
\end{equation}
In fact, after conversion to polar coordinates,
using $\frac{1}{\pi}\, du\,dv= dr^2 \, \frac{1}{2\pi}d\theta$,
the definition gives
\begin{equation}
\int_{\R^2} D\Phi \, F(\Phi^2)
= -\int_{\R^2}F'(r^2) \frac{1}{\pi} \, du \,  dv
 = -\int_0^\infty F'(t) dt = F(0),
\end{equation}
as claimed.
\end{example}

The supersymmetric version of the BFS--Dynkin isomorphism theorem
(see, for example, \cite[Corollary~11.3.7]{BBS-brief}),
relates random walks and superfields via an exact equality, as follows.

\begin{theorem}
\label{thm:BFS-Dynkin}
Let $F:\R^{N}\to \R$ be
such that $e^{\epsilon \sum_{z\in \Lambda}t_{z}} F (t)$ is a Schwartz
function for some $\epsilon >0$.  Then
\begin{equation}
\label{e:bfsdynkin}
    \int_0^\infty
    E_x\big(F(L_T) \1_{X(T)=y}\big) \, dT
    =
    \int_{\R^{2N}}
    D\Phi\, e^{-(\Phi,-\Delta\Phi)} F(\Phi^2) \,
    \bar\phi_x \phi_y
    ,
\end{equation}
where $\Delta$ is the generator (defined in \refeq{Deltadef})
of the random walk $X=(X(t))_{t\ge 0}$ with expectation $E_x$, and $L_T$ is the local time.
\end{theorem}

By definition, the two-point function \refeq{2ptfcndef} and expected length \refeq{EL0}
are given by expressions like the left-hand side of \refeq{bfsdynkin}, which
therefore can be rewritten as the right-hand side.

\subsubsection{Block-spin renormalisation}

The next lemma gives a way to rewrite the exponential factor on the right-hand side
of \refeq{bfsdynkin} as an integral over a single \emph{constant} block-spin
superfield $\Xinew=(\zeta,\bar\zeta,\xi,\bar\xi)$.
The application of this lemma can be regarded as a single block-spin renormalisation
group step, as in \cite[Section~1.4]{BBS-brief}.
For the statement of the lemma, we use the notation
\begin{equation}
  (\Xinew-\Phi,\Xinew-\Phi) = \sum_{x\in\Lambda}
    (\Xinew-\Phi_x)^2 = \sum_{x\in\Lambda}
  \Big((\zeta-\phi_x)(\zetab-\phib_x) + (\xi-\psi_x)(\xib-\psib_x)\Big).
\end{equation}

\begin{lemma} \label{lem:MF-decomp-super-new}
  For a superfield $\Phi = (\phi,\phib,\psi,\psib)=(\phi_x,\phib_x,\psi_x,\psib_x)_{x\in\Lambda}$,
  \begin{equation}
    \label{e:MF-decomp-super-new}
    e^{-(\Phi,-\Delta\Phi)}
    =
    \int_{\R^2}
    D\Xinew\,
    e^{-(\Xinew-\Phi,\Xinew-\Phi)}
    .
  \end{equation}
\end{lemma}

\begin{proof}
  Let $A\phi = N^{-1}\sum_{x\in\Lambda} \phi_x$.
  Since cross terms vanish,
    \begin{align}
    \sum_{x\in\Lambda} (\zeta-\phi_x)(\zetab-\phib_x)
    &=
    \sum_{x\in\Lambda} ((\zeta-A\phi)+(A\phi-\phi_x))((\zetab-A\phib)+(A\phib-\phib_x))
    \nnb& =
    N(\zeta-A\phi)(\zetab-A\phib) + \sum_{x\in \Lambda} (A\phi-\phi_x)(A\phib-\phib_x).
  \end{align}
  By definition of $\Delta$,
  and since $\sum_{x\in\Lambda}(\Delta f)_x=0$,
  the last term on the right-hand side is
  \begin{equation}
   \sum_{x\in \Lambda} (A\phi-\phi_x)(A\phib-\phib_x)
   =
   \sum_{x\in \Lambda}(A\phi-\phi_x)(\Delta\phib)_x = -\sum_{x\in\Lambda}\phi_x(\Delta\phib)_x.
   \end{equation}
   The fermionic part is completely analogous.
   Therefore,
   with $A\Phi = (A\phi,A\phib,A\psi,A\psib)$,
    \begin{align}
    (\Xinew-\Phi,\Xinew-\Phi)
    =
    N(\Xinew-A\Phi)^2 + (\Phi,-\Delta\Phi),
  \end{align}
  and hence
    \begin{align}
    \int_{\R^2} D\Xinew\, e^{-(\Xinew-\Phi,\Xinew-\Phi)}
    = e^{-(\Phi,-\Delta\Phi)}
    \int_{\R^2} D\Xinew\, e^{- N(\Xinew-A\Phi)^2}
    .
  \end{align}
  In the integral on the right-hand side, we make the change of variables
  $\zeta \mapsto \zeta+A\phi$, $\xi \mapsto \xi+A\psi$, and
  similarly for $\bar\zeta,\bar\xi$.  The bosonic change of variables is the
  usual one for Lebesgue integration, and the fermionic change of variables maintains the
  same $\bar\xi\xi$ term in $e^{- N(\Xinew-A\Phi)^2}$.  Thus the integral is unchanged and hence is equal to
  $\int_{\R^2} D\Xinew \, e^{- N\Xinew^2}$,
  which is $1$ by \refeq{intftau}.  This completes the proof.
\end{proof}

\subsection{Effective potential}

\begin{defn}
\label{def:V}
Given a (smooth) function $p:[0,\infty)\to [0,\infty)$ such that the following integral exists,
the \emph{effective potential} $V : [0,\infty) \to \R$ is defined by
\begin{equation}
  e^{-V(\Xinew^2)}
  =  \int_{\R^2} D\Phi \, e^{-(\Xinew-\Phi)^2} p(\Phi^2)
  .
\end{equation}
That the right-hand side truly is a function of the form $\Xinew^2$ is
proved in Lemma~\ref{lem:Vsusy}, which we defer to Section~\ref{sec:susylemmas}.
The consistency of this definition of $V$ with the formula given in \refeq{vintegral}
is established in Proposition~\ref{prop:Vint}.
\end{defn}

The proof of the next proposition appeals to Lemma~\ref{lem:Vsusy}.
Recall that $I_1$ is a modified Bessel function of the first kind.

\begin{prop}
\label{prop:Vint}
Fix  $\Xinew=(\zeta,\zetab,\xi,\xib)$.
For any bounded
smooth function $p:[0,\infty) \to [0,\infty)$ such that the integrals exist,
\begin{equation}
\lbeq{Vint}
    \int_{\R^2}
    D\Phi \, e^{-(\Xinew-\Phi)^2} p(\Phi^2)
    =
    e^{-\Xinew^2}
    \left(
    p(0) +
    \int_0^\infty p(s)e^{- s} \sqrt{\frac{\Xinew^2}{s}}I_1(2\sqrt{\Xinew^2  s}) ds
    \right),
\end{equation}
and hence
\begin{equation} \label{e:V-Bessel}
    V(t) = t - \log (p(0)+v(t)), \qquad
    v(t) = \int_0^\infty p(s)e^{- s} \sqrt{\frac{t}{s}}I_1(2\sqrt{ts}) ds.
\end{equation}
\end{prop}

\begin{proof}
We denote the left-hand side of \refeq{Vint} by $F=F(\zeta,\bar\zeta,\xi,\bar\xi)$.
By Lemma~\ref{lem:Vsusy}, $F$ is a function of $\Xinew^2$
so it suffices to prove that
\begin{equation}
    F(\zeta,\bar\zeta,0,0) =
    e^{-|\zeta|^2}
    \left(
    p(0) +
    \int_0^\infty p(s)e^{- s} |\zeta|\frac{1}{\sqrt{s}} I_1(2|\zeta|\sqrt{s}) ds
    \right).
\end{equation}
Let $\tilde{p}(s) = p(s) e^{- s}$.
With $\xi=\bar\xi=0$, the integrand of the left-hand side of \refeq{Vint} becomes
\begin{align}
    e^{- |\zeta|^2} e^{\zeta\phib + \bar\zeta\phi} \tilde{p}(\Phi^2)
    &=
    e^{- |\zeta|^2} e^{\zeta\phib + \zeta\phi} \left( \tilde{p}(|\phi|^2)
    + \tilde{p}'(|\phi|^2) \psi\psib \right).
\end{align}
Therefore, by the definition \refeq{intdef} of the integral,
and with
$I_0(z) = \frac{1}{2\pi}\int_0^{2\pi} e^{z\cos\theta} d\theta$
the modified Bessel function of the first kind,
\begin{align}
    F(\zeta,\bar\zeta,0,0) & =
    - e^{- |\zeta|^2}
    \int_{\R^2} e^{\zeta re^{-i\theta} + \bar\zeta re^{i\theta}}
    \tilde{p}'(r^2) \frac{ dr^2\, d\theta}{2\pi}
    \nnb & =
    - e^{- |\zeta|^2}
    \int_{0}^\infty ds\,
    \tilde{p}'(s)
    \int_0^{2\pi} e^{\zeta \sqrt{s} e^{-i\theta} + \bar\zeta \sqrt{s} e^{i\theta}}
    \frac{d\theta}{2\pi}
    \nnb & =
    - e^{- |\zeta|^2}
    \int_{0}^\infty ds \,
    \tilde{p}'(s)
    \int_0^{2\pi} e^{2|\zeta| \sqrt{s} \cos\theta}
    \frac{d\theta}{2\pi}
    \nnb & =
    - e^{- |\zeta|^2}
    \int_{0}^\infty
    \tilde{p}'(s) I_0(2|\zeta|\sqrt{s}) \, ds
    \nnb & =
     e^{- |\zeta|^2}
     \left(\tilde{p}(0)+
    \int_{0}^\infty
    \tilde{p}(s) I_0'(2|\zeta|\sqrt{s})  |\zeta|\frac{1}{\sqrt{s}} ds\right),
\end{align}
where we used integration by parts for the last equality, together with our
assumption that $p$ is bounded at infinity (this can certainly be weakened).
Since $I_0'=I_1$, the proof is complete.
\end{proof}

Next, for later use, we state and prove a lemma that shows how the effective
potential arises in various integrals.  As usual, we write
$V' = \frac{dV}{dt}$ and
$\dot{V} = \ddp{V}{\epsilon}|_{\epsilon=0}$ (with the $\epsilon$-dependence as
in \refeq{ELdot}, see \refeq{Vdotdef}).  We also write $\Vpm'=1-V'$.
We define forms $k_*=k_*(\Xinew^2,\zeta,\zetab)$ by:
\begin{equation}
\begin{gathered}
    k_0  = \zeta Q'(\Xinew^2), \qquad \bar{k}_0 = \zetab Q'(\Xinew^2),
    \qquad k_{00} = Q'(\Xinew^2) + Q'(\Xinew^2)^2|\zeta|^2 - V''(\Xinew^2)|\zeta|^2,
    \\
    k_\taunew  = \Vdot(\Xinew^2), \qquad
    k_{0\taunew}  = \zeta(Q'(\Xinew^2)\Vdot(\Xinew^2) + \Vdot'(\Xinew^2)),
    \qquad
    \bar k_{0\taunew}  = \zetab (Q'(\Xinew^2)\Vdot(\Xinew^2) + \Vdot'(\Xinew^2)),
    \\
    k_{00\taunew}  = k_{00}\Vdot(\Xinew^2)
    +
    \big( 1 + 2\Vpm'(\Xinew^2) |\zeta|^2 \big) \dot{V}'(\Xinew^2)
    +
    \dot{V}''(\Xinew^2)|\zeta|^2
    .
\lbeq{kdefs}
\end{gathered}
\end{equation}

\begin{lemma}
\label{lem:integrals}
The following integral formulas hold:
\begin{equation}
\begin{gathered}
\lbeq{phi}
     \int_{\R^2} D\Phi\, \phi  p(\Phi^2) e^{-(\Xinew-\Phi)^2}
     =
    k_0
    e^{-V(\Xinew^2)},
    \qquad
    \int_{\R^2} D\Phi\, \phib p(\Phi^2) e^{-(\Xinew-\Phi)^2}
     =
    \bar{k}_0
    e^{-V(\Xinew^2)},
    \\
    \int_{\R^2} D\Phi\,  \phib \phi  p(\Phi^2) e^{-(\Xinew-\Phi)^2}
     =
    k_{00} e^{-V(\Xinew^2)},
    \qquad
    \int_{\R^2} D\Phi\,  \Phi^2 p(\Phi^2)e^{-(\Xinew-\Phi)^2}
     = k_\taunew e^{-V(\Xinew^2)},
    \\
    \int_{\R^2} D\Phi\,  \phi \Phi^2 p(\Phi^2) e^{-(\Xinew-\Phi)^2}
     =
    k_{0\taunew} e^{-V(\Xinew^2)},
    \qquad
    \int_{\R^2}D\Phi\,  \phib \Phi^2 p(\Phi^2) e^{-(\Xinew-\Phi)^2}
     =
    \bar k_{0\taunew} e^{-V(\Xinew^2)},
    \\
    \int_{\R^2}D\Phi\,  \phib \phi \Phi^2 p(\Phi^2) e^{-(\Xinew-\Phi)^2}
     =
    k_{00\taunew}
    e^{-V(\Xinew^2)}.
\end{gathered}
\end{equation}
\end{lemma}

\begin{proof}
Given $h:\Lambda \to \C$,
let $(\Xinew+h)^2= (\zeta+h)(\bar\zeta + \bar h) + \xi\bar\xi$.
There is no fermionic partner for $h$ in $(\Xinew+h)^2$.
Completion of the square and the definition of $V$ give
\begin{align}
\lbeq{hintegral}
  \int_{\R^2} D\Phi\, e^{- (\Xinew-\Phi)^2} p(\Phi^2) e^{h\phib + \bar h \phi}
  =
  e^{ h \bar h} e^{ h\zetab +  \bar h \zeta} \int_{\R^2} D\Phi\, e^{- (\Xinew-\Phi+h)^2} p(\Phi^2)
  =
  e^{ h\bar h} e^{ h\zetab +  \bar h \zeta} e^{-V((\Xinew+h)^2)}
  .
\end{align}
Therefore, using $\ddp{\bar h}{h}=0$, we obtain
\begin{align}
    \int_{\R^2} D\Phi\, \phib p(\Phi^2) e^{-(\Xinew-\Phi)^2}
    & =
    \ddp{}{h}\Big|_{h=0}
    \int_{\R^2} D\Phi\, e^{- (\Xinew-\Phi)^2} p(\Phi^2) e^{h\phib + \bar h \phi}
    \nnb & =
    \ddp{}{h}\Big|_{h=0} e^{ h\bar h} e^{ h\zetab +  \bar h \zeta} e^{-V((\Xinew+h)^2)}
    =
    \zetab (1-V'(\Xinew^2))e^{-V(\Xinew^2)}.
\end{align}
The first and third equalities in \refeq{phi} follow similarly.
For the fourth, with $V^{(\epsilon)}$ the effective potential for $p(s)e^{-\epsilon s}$,
we use
\begin{align}
    \int_{\R^2} D\Phi \, \Phi^2 p(\Phi^2)e^{-(\Xinew-\Phi)^2}
    &=
    - \ddp{}{\epsilon} \Big|_{\epsilon=0}
    \int_{\R^2} D\Phi\, p(\Phi^2)e^{-\epsilon \Phi^2}e^{-(\Xinew-\Phi)^2}
    \nnb & =
    - \ddp{}{\epsilon} \Big|_{\epsilon=0} e^{-V^{(\epsilon)}(\Xinew^2)}
    = \dot{V}(\Xinew^2) e^{-V(\Xinew^2)}.
\end{align}
The remaining three identities follow, e.g., from
\begin{align}
    \int_{\R^2} D\Phi\, \phib \Phi^2 p(\Phi^2) e^{-(\Xinew-\Phi)^2}
    & =
    - \ddp{}{\epsilon}\Big|_{\epsilon=0} \int_{\R^2} D\Phi \, \phib  p(\Phi^2)
    e^{-\epsilon \Phi^2} e^{-(\Xinew-\Phi)^2},
\end{align}
together with differentiation
of the right-hand sides of the first three identities with respect to $\epsilon$.
\end{proof}

\subsection{Two-point function and expected length}

We now have what is needed
to prove integral representations for the two-point function
and expected length, in the next two propositions.

\begin{prop}
\label{prop:2ptfcn}
The two-point function is given by
\begin{align}
    G_{01} & =
     \int_{\R^2} D\Xinew \, e^{-NV(\Xinew^2)}  (1-V'(\Xinew^2))^2 |\zeta|^2  ,
\\
    G_{00} & =  \
    (1-V'(0)) + G_{01} -
    \int_{\R^2} D\Xinew\, e^{-NV(\Xinew^2)}  V''(\Xinew^2)|\zeta|^2 .
\lbeq{G00int}
\end{align}
\end{prop}

\begin{proof}
By the the definition of the two-point function in \refeq{2ptfcndef},
followed by the supersymmetric BFS--Dynkin isomorphism~\refeq{bfsdynkin}
and the block-spin transformation of Lemma~\ref{lem:MF-decomp-super-new},
\begin{align}
    G_{xy}
    & =
    \int_0^\infty E_x( p_N(L_T) \1_{X(T)=y}) \, dT
    \nnb & =
    \int_{\R^{2N}} D\Phi\, e^{-(\Phi,-\Delta\Phi)}\phib_x \phi_y
    p_N(\Phi^2)
    \nnb & =
    \label{e:sawrep1}
    \int_{\R^2} D\Xinew\, \int_{\R^{2N}} D\Phi\, \phib_x\phi_y   p_N(\Phi^2) e^{-(\Xinew-\Phi)^2}.
\end{align}
The integral over $\R^{2N}$ on the right-hand side of \refeq{sawrep1} factorises
into a product of $N$ integrals over $\R^2$ (each an integral with respect to $\Phi_x$
at a single point $x$).
With the definition of the effective potential in Definition~\ref{def:V},
and with the first line of \refeq{phi}, this
leads to
\begin{align}
    G_{01} & =  \int_{\R^2} D\Xinew \, e^{-(N-2)V(\Xinew^2)}
    \left( \int_{\R^2} D\Phi_0\,  \phib_0 p(\Phi_0^2) e^{-(\Xinew-\Phi_0)^2}\right)
    \left( \int_{\R^2} D\Phi_1\, \phi_1 p(\Phi_1^2) e^{-(\Xinew-\Phi_1)^2}\right)
    \nnb & =
     \int_{\R^2} D\Xinew\, e^{-NV(\Xinew^2)}  (1-V'(\Xinew^2))^2 |\zeta|^2  .
\lbeq{G01int}
\end{align}
Similarly, by the third equality of \refeq{phi},
\begin{align}
    G_{00} & =  \int_{\R^2} D\Xinew\, e^{-(N-1)V(\Xinew^2)}
    \int_{\R^2} D\Phi_0\,  \phib_0\phi_0 p(\Phi_0^2) e^{-(\Xinew-\Phi_0)^2}
    \nnb & =
    \int_{\R^2} D\Xinew\, e^{-NV(\Xinew^2)}
    \big( (1-V'(\Xinew^2)) +(1-V'(\Xinew^2))^2 |\zeta|^2
    - V''(\Xinew^2)|\zeta|^2 \big)
    \nnb & =
    (1-V'(0)) + G_{01} -
    \int_{\R^2} D\Xinew\, e^{-NV(\Xinew^2)}  V''(\Xinew^2)|\zeta|^2 ,
\lbeq{G00intpf}
\end{align}
where in the last line
we used \refeq{intftau} for the first term and \refeq{G01int} for the second.
This completes the proof.
\end{proof}

For the expected length the general procedure is the same.
With $k_*$ defined by \refeq{kdefs}, let
\begin{align}
\lbeq{K0xydef}
    K_{0xy}
    & =
    \begin{cases}
    \bar k_0 k_0 k_{\taunew} & (x=1,\, y=2)
    \\
    k_{00}
    k_{\taunew}
    & (x=0,\, y=1)
    \\
    \bar k_0 k_{0\taunew}
    & (x=y=1)
    \\
       k_{00\taunew} & (x=y=0)
    .
    \end{cases}
\end{align}

\begin{prop}
\label{prop:EL}
The expected length is given by
\begin{align}
    \Ex L
    & =
    \frac{1}{\chi} \Big(
    (N-1)(N-2)\int_{\R^2} D\Xinew\, e^{-NV(\Xinew^2)}K_{012}
    \nnb & \quad\quad + (N-1) \int_{\R^2} D\Xinew\, e^{-NV(\Xinew^2)}(K_{001}+ 2K_{011})
    + \int_{\R^2}D\Xinew\,e^{-NV(\Xinew^2)}K_{000} \Big).
\end{align}
\end{prop}

\begin{proof}
By \refeq{EL0} and $T=\sum_{x\in \Lambda}L_{T,x}$,
\begin{align}
    \Ex L
    & =
    \frac{1}{\chi} \sum_{x,y\in\Lambda} \int_0^\infty  E_0(L_{T,y} p_N(L_T)\1_{X(T)=x}) dT.
\end{align}
By the supersymmetric BFS--Dynkin isomorphism~\refeq{bfsdynkin}
followed by the block-spin transformation of Lemma~\ref{lem:MF-decomp-super-new},
\begin{align}
    \Ex L
    & =\frac{1}{\chi} \sum_{x,y\in\Lambda}
    \int_{\R^{2N}} D\Phi\, \phib_0\phi_x \Phi^2_y p_N(\Phi^2)
    e^{-(\Phi,-\Delta\Phi)}
    \nnb & =
    \frac{1}{\chi} \sum_{x,y\in\Lambda}
    \int_{\R^2} D\Xinew\,\int_{\R^{2N}}D\Phi\, \phib_0\phi_x \Phi^2_y p_N(\Phi^2)
    e^{-(\Xinew-\Phi)^2}.
\end{align}
By symmetry, it suffices to show that
\begin{equation}
\lbeq{K0xyint}
    \int_{\R^{2N}}D\Phi\, \phib_0\phi_x \Phi^2_y p(\Phi^2) e^{-(\Xinew-\Phi)^2}
    =
    e^{-NV(\Xinew^2)}K_{0xy}.
\end{equation}
For the case of $0,x,y$ distinct, since the $N-3$ integrals
for the factors with $z\neq 0,x,y$ are the same,
\begin{multline}
  \int_{\R^{2N}}D\Phi\, \phib_0\phi_x \Phi^2_y p(\Phi^2) e^{-(\Xinew-\Phi)^2}
  =
    e^{-(N-3)V(\Xinew^2)}
    \left( \int_{\R^2} D\Phi_0\, \phib_0 p(\Phi_0^2)e^{-(\Xinew-\Phi_0)^2} \right) \\
    \times
    \left( \int_{\R^2} D\Phi_x\, \phi_x p(\Phi_x^2)e^{-(\Xinew-\Phi_x)^2} \right)
    \left( \int_{\R^2} D\Phi_y\, \Phi^2_y p(\Phi_y^2)e^{-(\Xinew-\Phi_y)^2} \right)
    ,
\end{multline}
and \refeq{K0xyint} follows from the definitions
of $k_0, \bar{k}_0, k_\taunew$. The other cases are similar.
\end{proof}

\subsection{Supersymmetry}
\label{sec:susylemmas}

In this section, we prove Lemma~\ref{lem:Vsusy}, which was used in the
proof of Proposition~\ref{prop:Vint}.  It is possible to give a more direct proof
of Proposition~\ref{prop:Vint} without using the notion of supersymmetry.  However,
the proof using Lemma~\ref{lem:Vsusy} is particularly elegant.

The \emph{supersymmetry generator} is the anti-derivation defined by
\begin{equation}
\label{e:Qdef}
    Q
    =
    \psi \frac{\partial}{\partial \phi} +
    \bar\psi \frac{\partial}{\partial \bar\phi}-
    \phi \frac{\partial}{\partial \psi}  +
    \bar\phi \frac{\partial}{\partial \bar\psi}
    .
\end{equation}
We say that $F=F(\phi,\phib,\psi,\psib)$ is \emph{supersymmetric} if $QF=0$.
The next lemma is \cite[Lemma~A.4]{BI03d}.

\begin{lemma}
\label{lem:evenSUSY}
If $F$ is even and supersymmetric then $F=f(\Phi^2)$ for some function $f$.
\end{lemma}

\begin{proof}
We write
$F = G  + H\psi\psib$, and use subscripts to denote partial derivatives.
It suffices to show that there is a function $f$ such that $G(\phi,\phib)=f(|\phi|^2)$
and $H(\phi,\phib) = f'(|\phi|^2)$.
Since
\begin{align}
  0=QF & =
  G_\phi \psi  + G_{\phib} \psib
  -\phi H\psib - \phib H \psi
  ,
\end{align}
we see that $G_\phi = \phib H$ and $G_{\phib}=\phi H$.
Therefore,
\begin{align}
    \frac{d}{d\theta}G(\phi e^{i\theta}, \phib e^{-i\theta})
    & = G_\phi(\phi e^{i\theta}, \phib e^{-i\theta}) \phi i e^{i\theta}
    + G_{\phib}(\phi e^{i\theta}, \phib e^{-i\theta}) \phib (-i) e^{-i\theta}
    \nnb
    & =
    \phib e^{-i\theta} H(\phi e^{i\theta}, \phib e^{-i\theta}) \phi i e^{i\theta}
    + \phi e^{i\theta} H(\phi e^{i\theta}, \phib e^{-i\theta}) \phib (- i) e^{-i\theta}
    =0
    .
\end{align}
 This implies that
there is a function $f$ as required.
\end{proof}

\begin{lemma}
\label{lem:Vsusy}
The integral $\int_{\R^2} D\Phi \, e^{-(\Xinew-\Phi)^2}p(\Phi^2)$ is an even supersymmetric
form, and hence is a function of $\Xinew^2$.
\end{lemma}

\begin{proof}
Let $F=F(\zeta,\zetab,\xi,\xib)=\int_{\R^2} D\Phi \, e^{-(\Xinew-\Phi)^2}p(\Phi^2)$.
Since $p(\Phi^2)$ is even, only even contributions in $\psi,\psib$ from
\begin{equation}
    e^{-(\Xinew-\Phi)^2} = e^{-|\zeta-\phi|^2}(1 - (\xi-\psi)(\xib-\psib))
\end{equation}
can contribute to the integral.  Thus, within the integral,
the above right-hand side can be replaced
by $e^{-|\zeta-\phi|^2}(1-\xi\xib-\psi\psib)$, and we see that $F$ is even in $\xi,\xib$.

To see that $F$ is supersymmetric,
let $Q_\Xinew$ act on $\Xinew=(\zeta,\zetab,\xi,\xib)$ and $Q_\Phi$ on $\Phi=(\phi,\phib,\psi,\psib)$.
By definition,
\begin{equation}
    e^{-(\Xinew-\Phi)^2} = e^{-\Phi^2}e^{-\Xinew^2} e^K
    \quad\text{with}\quad
    K = \zeta\phib+\phi\zetab +\xi\psib+\psi\xib.
\end{equation}
Let $\tilde p(\Phi^2) = p(\Phi^2)e^{-\Phi^2}$.
Since $Q_\Xinew$ is an anti-derivation,  since $Q_\Xinew e^{-\Xinew^2}=0$
and $Q_\Phi \tilde p(\Phi^2)=0$
(by \cite[Example~11.4.4]{BBS-brief}),
and since $Q_\Xinew e^K = - Q_\Phi e^K$,
\begin{align}
    Q_\Xinew F
    & = e^{-\Xinew^2} Q_\Xinew \Big( \int_{\R^2} D\Phi\, e^K \tilde p(\Phi^2)\Big)
    = e^{-\Xinew^2}  \int_{\R^2} D\Phi\, (Q_\Xinew e^K )\tilde p(\Phi^2)
    \nnb & =
    e^{-\Xinew^2}  \int_{\R^2} D\Phi\,  (-Q_\Phi e^K )\tilde p(\Phi^2)
    =
    -e^{-\Xinew^2} \int_{\R^2} D\Phi\,  Q_\Phi \Big( e^K \tilde p(\Phi^2) \Big) .
\end{align}
The last integrand is in the image of $Q_{\Phi}$, so the integral is zero
(see \cite[Section~11.4.1]{BBS-brief}), and hence $F$ is supersymmetric.
By Lemma~\ref{lem:evenSUSY}, $F$ is therefore a function of $\Xinew^2$.
\end{proof}

\section{Proof of general results:
  Theorems~\ref{thm:2ptfcn-mr}--\ref{thm:mr}}
\label{sec:pf-mr}

The proofs of Theorems~\ref{thm:2ptfcn-mr}--\ref{thm:mr} amount to application of the Laplace method
to the integrals of Propositions~\ref{prop:2ptfcn}--\ref{prop:EL}.
The application of the Laplace method depends on whether: (i) the global minimum of the
effective potential is attained at zero and only at zero, or (ii)
it is attained at a point $t_0>0$ with $V(t_0)<0$ or $V(t_0)=0$.
Case~(i) concerns  the
dilute phase, the second-order curve, and the tricritical point,
while case~(ii) concerns the dense phase and first-order curve.

\subsection{Laplace method}

\subsubsection{Laplace method: minimum at endpoint}

For the dilute phase, the second-order curve, and the tricritical point,
we use the following theorem, which can be found, e.g., in \cite[p.81]{Olve97}.
The theorem can be extended to an asymptotic expansion to all orders, \cite[p.86]{Olve97}
or \cite[p.233]{Olve70},
but we do not need the extension.  In a corollary to the theorem,
we adapt its statement to integrals of the form appearing in Propositions~\ref{prop:2ptfcn}--\ref{prop:EL}.

\begin{theorem}
\label{thm:Laplace-endpoint}
Suppose that $V,\SD: [a,b) \to \R$ ($b=\infty$ is allowed)
are such that: \\
(i)
$V$ has a unique global minimum $V(a)$ (as defined at the beginning of Section~\ref{sec:results}),
\\
(ii) $V'$ and $\SD$ are continuous in a neighbourhood of $a$, except possibly at $a$, \\
(iii) as $t \to a^+$,
$V(t) \sim v_0(t-a)^\mu$,
$V'(t) \sim \mu v_0(t-a)^{\mu-1}$,
$\SD(t) \sim q_0(t-a)^{\lambda-1}$,
with $v_0,\mu,\lambda>0$ and $q_0\neq 0$,
\\
(iv) $e^{-NV(t)}\SD(t)$ is integrable for large $N$. \\
Then
\begin{equation}
\lbeq{Laplace-endpoint}
    \int_a^b e^{-NV(t)} q(t) dt
    \sim
    e^{-NV(a)}
    \frac{q_0}{\mu (v_0
    N)^{\lambda/\mu}}
    \Gamma \Big( \frac{\lambda}{\mu} \Big)
     .
\end{equation}
\end{theorem}

\begin{cor}
\label{cor:Laplace-endpoint}
Suppose that the hypotheses on $V$ of Theorem~\ref{thm:Laplace-endpoint} hold
with $a=0$ and $b=\infty$.  Suppose that
$F:[0,\infty)^2 \to \R$ is such that $\SD(t)=V'(t)F(t,t)$ and $\SD(t)=\partial_1 F(t,t)$
obey hypotheses (ii) and (iv) of Theorem~\ref{thm:Laplace-endpoint},
and that, as $t \downarrow 0$,
\begin{align}
\lbeq{FdF}
    F(t,t) & \sim  q_0 t^{\lambda_0},
    \qquad
    \partial_1 F(t,t)  \sim \lambda_1 r_0 t^{\lambda_1-1}.
\end{align}
If $\lambda_1 > \lambda_0 \ge 0$ then
\begin{align}
    \int_{\R^2} D\Xinew\, e^{-NV(\Xinew^2)}F(\Xinew^2, |\zeta|^2)
    & \sim
    e^{-NV(0)} \frac{q_0}{(v_0
    N)^{\lambda_0/\mu}}
    \Gamma\left( \frac{\mu+\lambda_0}{\mu}\right)
     .
\lbeq{CRint}
\end{align}
If $\lambda_1=\lambda_0>0$ then
\begin{align}
\lbeq{CRint1}
    \int_{\R^2} D\Xinew\, e^{-NV(\Xinew^2)}F(\Xinew^2, |\zeta|^2)
    & \sim
    e^{-NV(0)}
    \frac{q_0- r_0}{(v_0
    N)^{\lambda_0/\mu}}
    \Gamma\left( \frac{\mu+\lambda_0}{\mu}\right)
     ,
\end{align}
where the right-hand side is interpreted as $e^{-NV(0)}o(N^{-\lambda_0/\mu})$ if $q_0=r_0$.
\\
For any $\lambda_1>0$ and $\lambda_0 \ge 0$,
the integral is at most $e^{-NV(0)}O(N^{-\lambda_0/\mu}+ N^{-\lambda_1/\mu})$.
\end{cor}

\begin{proof}
By definition of the integral, and
since $\bar\xi\xi = \frac{1}{\pi}dxdy = \frac{1}{2\pi}dr^2d\theta$
(as in \refeq{intftau}),
\begin{align}
    &\int_{\R^2} D\Xinew\, e^{-NV(\Xinew^2)}F(\Xinew^2, |\zeta|^2)
    \nnb & \quad =
    \int_{\R^2} D\Xinew\, e^{-NV(|\zeta|^2)}\Big(1-NV'(|\zeta|^2)\xi\bar\xi\Big)\Big(F(|\zeta|^2,|\zeta|^2) + \partial_1 F(|\zeta|^2,|\zeta|^2)\xi\bar\xi\Big)
    \nnb &\quad =
    \int_{\R^2} D\Xinew\, e^{-NV(|\zeta|^2)}
    \Big(NV'(|\zeta|^2)F(|\zeta|^2,|\zeta|^2) - \partial_1 F(|\zeta|^2,|\zeta|^2)\Big) \bar\xi\xi
    \nnb &\quad =
    \int_0^\infty e^{-NV(t)}(NV'(t)F(t,t) - \partial_1 F(t,t)) \, dt.
\lbeq{CRintpf}
\end{align}
Now we apply Theorem~\ref{thm:Laplace-endpoint}.
Since $V'(t)F(t,t) \sim \mu v_0
t^{\mu-1}q_0t^{\lambda_0}$,
the power of $N$ arising for this term is $N N^{-(\mu+\lambda_0)/\mu}
= N^{-\lambda_0/\mu}$.  If $\lambda_1 > \lambda_0 \ge 0$ then
this dominates the power $N^{-\lambda_1/\mu}$ from the $\partial_1
F(t,t)$ term and yields \refeq{CRint}.
If $\lambda_1=\lambda_0 >0$ then both terms contribute the same power of $N$,
and \refeq{CRint1} follows from
$\Gamma((\mu+\lambda_1)/\mu)=(\lambda_1/\mu)\Gamma(\lambda_1/\mu)$.
Finally, the general upper bound follows immediately from the above considerations.
\end{proof}

\subsubsection{Laplace method: minimum at interior point}

The following theorem from \cite[p.127]{Olve97} more than covers our needs for the case
where $V$ attains
its unique global minimum in an open interval.
Its analyticity assumption could be weakened, but the analyticity
does hold in our setting.

\begin{theorem}
\label{thm:Laplace-interior}
Let $a\in [-\infty,\infty)$ and $b\in (-\infty,\infty]$.
Suppose that $V,\SD: (a,b)\to\R$ are analytic,
and that $V$ has a unique global minimum at $t_0\in (a,b)$
(as defined at the
beginning of Section~\ref{sec:results})
with $V'(t_0)=0$ and $V''(t_0)>0$.  Then, assuming that the integral is finite
for some $N$,
\begin{equation}
\lbeq{Laplace-interior}
    \int_a^b e^{-NV(t)} \SD(t) dt
    \sim
    2e^{-NV(t_0)}
    \sum_{s=0}^\infty
    \Gamma (s+1/2) \frac{b_{s}}{N^{s+1/2}} ,
\end{equation}
with (all functions evaluated at $t_0$)
\begin{align}
\lbeq{b0}
    b_0 & = \frac{\SD}{(2 V'')^{1/2}},
    \\
\lbeq{b1}
    b_1 & =
    \left(
    2\SD'' - \frac{2V'''\SD'}{V''}
    +
    \left[
    \frac{5V'''^2}{6V''^2} -\frac{V''''}{2V''}
    \right] \SD
    \right)
    \frac{1}{(2V'')^{3/2}}
    ,
\end{align}
and with $b_{s}$ as given in \cite{Olve97} for $s \ge 2$.
\end{theorem}

\begin{cor}
\label{cor:Laplace-interior}
Suppose that $V: (0,\infty) \to\R$ is analytic
and has a unique global minimum at $t_0\in (0,\infty)$
(as defined at the beginning of Section~\ref{sec:results})
with $V'(t_0)=0$ and $V''(t_0)>0$.
Given $F: (0,\infty)^2 \to \R$, suppose that the functions
$C(t)=V'(t)F(t,t)$ and $D(t)=\partial_1F(t,t)$ are analytic on $(a,b)$
and that $\int_0^\infty e^{-NV(t)} \SD(t) \, dt$ is finite for $q=C$ and $q=D$.
Then
\begin{align}
    \int_{\R^2} D\Xinew\, e^{-NV(\Xinew^2)} F(\Xinew^2,|\zeta|^2) & \sim
    2e^{-NV(t_0)}
    \sum_{s=0}^\infty
    \frac{1}{N^{s+1/2}}
    \Big( c_{s+1}\Gamma ( s+3/2)   -d_s\Gamma ( s+1/2 ) \Big) ,
\end{align}
where the coefficients $c_s$ and $d_s$ are the coefficients
$b_s$ computed when the function $\SD$
in Theorem~\ref{thm:Laplace-interior} is replaced
by $C$ and $D$, respectively.
Assuming that $\partial_2 F(t_0,t_0)\neq 0$, we have in particular
\begin{align}
\lbeq{s0}
    \int_{\R^2} D\Xinew\, e^{-NV(\Xinew^2)} F(\Xinew^2,|\zeta|^2) & \sim
    e^{-NV(t_0)} \frac{1}{N^{1/2}}
    \frac{\sqrt{2\pi}}{V''(t_0)^{1/2}}
    \partial_2 F(t_0,t_0)
    .
\end{align}
\end{cor}

\begin{proof}
The full expansion follows from Theorem~\ref{thm:Laplace-interior} and \refeq{CRintpf}.
Let $A  = \frac{\sqrt{2}}{V''(t_0)^{1/2}} \partial_2 F(t_0,t_0)$.
For \refeq{s0},
since $\Gamma(1/2)=\sqrt{\pi}$, it suffices to show that $c_1-2d_0=A$.
Let $F' = \frac{d}{dt}F(t,t) = \partial_1F(t,t)+\partial_2F(t,t)$.
Then
\begin{align}
    C' & = V'' F + V' F', \qquad C'' = V''' F + 2V''F' + V'F''.
\end{align}
Since $V'(t_0)=0$ we find from \refeq{b0}--\refeq{b1} that
\begin{align}
    c_1-2d_0 & = \frac{1}{(2V'')^{3/2}} \left(2C'' - \frac{2V'''C'}{V''} \right)
    - 2 \frac{\partial_1 F}{(2V'')^{1/2}}
    \nnb & =
    \frac{1}{(2V'')^{3/2}} \left(2(V''' F + 2V''F') - \frac{2V'''V'' F}{V''} \right)
    - 2 \frac{\partial_1 F}{(2V'')^{1/2}},
\end{align}
and after simplification the right-hand side is equal to $A$.
\end{proof}

On the first-order curve, $V$ has global minima $V(0)=V(t_0)$ with $V'(0)>0$ and
$V''(t_0)>0$ (by smoothness of $V$, also $V'(t_0)=0$).
The following corollary covers the cases we need.

\begin{cor}
\label{cor:Laplace-2min}
Suppose that $V: (0,\infty) \to\R$ is analytic
and has global minima $V(0)=V(t_0)=0$ for $t_0\in (0,\infty)$
(as defined at the beginning of Section~\ref{sec:results})
with $V'(0)>0$, $V'(t_0)=0$, and $V''(t_0)>0$.
With the notation of Corollary~\ref{cor:Laplace-endpoint},
assume that $\lambda_1\ge\lambda_0\ge 1$,
and with the notation of Corollary~\ref{cor:Laplace-interior},
assume that $\partial_2 F(t_0,t_0)\neq 0$. Then, for $F$ obeying the
assumptions of Corollaries~\ref{cor:Laplace-endpoint} and \ref{cor:Laplace-interior},
\begin{align}
\lbeq{2min}
    \int_{\R^2} D\Xinew\, e^{-NV(\Xinew^2)} F(\Xinew^2,|\zeta|^2) & \sim
     \frac{1}{N^{1/2}}
    \frac{\sqrt{2\pi}}{V''(t_0)^{1/2}}
    \partial_2 F(t_0,t_0)
    .
\end{align}
If instead $\lambda_1 > \lambda_0=0$, then the right-hand side of \refeq{2min}
is at most $O(1)$.
\end{cor}

\begin{proof}
By \refeq{CRintpf},
\begin{align}
    \int_{\R^2} D\Xinew\, e^{-NV(\Xinew^2)}F(\Xinew^2, |\zeta|^2)
    &  =
    \int_0^\infty e^{-NV(t)}(NV'(t)F(t,t) - \partial_1 F(t,t)) \, dt.
\lbeq{CRintpf-2min}
\end{align}
We divide the integral on the right-hand side into integrals over $(0,\frac 12 t_0)$
and $(\frac 12 t_0,\infty)$.
Exactly as in the proof of Corollary~\ref{cor:Laplace-endpoint} with $\mu=1$ (only changing the integration interval),
if $\lambda_1\ge\lambda_0\ge 1$ then the former integral
is at most $O(N^{-1})$.
Exactly as in the proof of
Corollary~\ref{cor:Laplace-interior}, the latter integral is
asymptotic to the right-hand side of \refeq{2min}, which dominates $O(N^{-1})$.
If instead $\lambda_0=0$ then by Corollary~\ref{cor:Laplace-endpoint} there can
be a contribution from $t=0$ which is $O(1)$.
\end{proof}

\subsection{Two-point function and susceptibility}

We now prove Theorem~\ref{thm:2ptfcn-mr} and the part of Theorem~\ref{thm:mr} that
concerns the susceptibility.
For convenience, we restate Theorem~\ref{thm:2ptfcn-mr} as the following proposition.

\begin{prop}
\label{prop:2ptfcn-asy}
The two-point function has the asymptotic behaviour:
\begin{align}
\lbeq{G01asy}
    G_{01} & \sim
    \begin{cases}
    \frac{(1-V'(0))^2}{V'(0)N}\Gamma(2/1) & \qquad\qquad\text{(dilute phase)}
    \\
    \frac{1}{(\frac {1}{2!} V''(0)N)^{1/2}} \Gamma(3/2) & \qquad\qquad\text{(second-order curve)}
    \\
    \frac{1}{(\frac {1}{3!} V'''(0)N)^{1/3}} \Gamma(4/3) & \qquad\qquad\text{(tricritical point)}
    \\
    \frac{e^{-NV(t_0)}}{N^{1/2}} \frac{\sqrt{2\pi}}{V''(t_0)^{1/2}}
    & \qquad\qquad\text{(dense phase and first-order curve)},
    \end{cases}
\\
\lbeq{G00asy}
    G_{00} & \sim
    \begin{cases}
    1-V'(0) & \text{(dilute phase and first-order curve)}
    \\
    1 & \text{(second-order curve)}
    \\
    1 & \text{(tricritical point)}
    \\
    \frac{e^{-NV(t_0)}}{N^{1/2}} \frac{\sqrt{2\pi}}{V''(t_0)^{1/2}}(1-V''(t_0))
    & \text{(dense phase)}.
    \end{cases}
\end{align}
\end{prop}

\begin{proof}
By Proposition~\ref{prop:2ptfcn},
\begin{align}
    G_{01} & =
     \int_{\R^2} D\Xinew \, e^{-NV(\Xinew^2)}  (1-V'(\Xinew^2))^2 |\zeta|^2
     .
\lbeq{G01bis}
\end{align}
(Note that \refeq{G01example} then follows via \refeq{CRintpf}.)
The integrability of \refeq{G01bis} follows from
the lower bound on $V$ of Proposition~\ref{prop:Vderivs0}(i).
For the first three cases of \refeq{G01asy}, we apply  Corollary~\ref{cor:Laplace-endpoint} with
\begin{align}
    &\mu=1, \quad v_0 = V'(0) &\text{(dilute phase)}
    \\
    &\mu=2, \quad v_0 = \frac{1}{2!}V''(0)  &\text{(second-order curve)}
     \\
    &\mu=3, \quad v_0 = \frac{1}{3!}V'''(0)  &\text{(tricritical point).}
\end{align}
The integrand of \refeq{G01bis} involves $F_{01}(\Xinew^2,|\zeta|^2) = (1-V'(\Xinew^2))^2|\zeta|^2$, for which
\begin{equation}
\lbeq{F01tt}
    F_{01}(t,t) = (1-V'(t))^2 t,
    \qquad
    \partial_1 F_{01}(t,t) = 2(1-V'(t))(-V''(t)) t
    .
\end{equation}
From this, we see that
\begin{align}
    &\lambda_0=1, \quad\lambda_1 \ge 2,
    \quad q_0=(1-V'(0))^2 &\text{(dilute phase and second-order curve)}
     \\
    &\lambda_0=1, \quad \lambda_1=3, \quad q_0 = 1  &\text{(tricritical point).}
\end{align}
In all three cases $\lambda_1>\lambda_0$, so Corollary~\ref{cor:Laplace-endpoint} gives,
as desired,
\begin{equation}
    G_{01} \sim
    \frac{q_0}{(v_0N)^{\lambda_0/\mu}}
    \Gamma\left( \frac{\mu+\lambda_0}{\mu} \right)
\end{equation}
(recall that $V'(0)=0$ on the second-order curve).

For the first three cases of \refeq{G00asy}, by Proposition~\ref{prop:2ptfcn},
\begin{align}
\lbeq{G00sum}
    G_{00} & =
    (1-V'(0)) + G_{01} -
    \int_{\R^2} D\Xinew\, e^{-NV(\Xinew^2)}  V''(\Xinew^2)|\zeta|^2 .
\end{align}
The integral has $F_{00}(\Xinew^2,|\zeta|^2) = V''(\Xinew^2)|\zeta|^2$, so
\begin{equation}
    F_{00}(t,t) = V''(t)t, \qquad \partial_1 F_{00}(t,t) = V'''(t)t
    .
\end{equation}
The dilute, second-order, and tricritical cases have respectively:
$\lambda_0 \ge 1$, $\lambda_1 \ge 2$; $\lambda_0=1$, $\lambda_1 \ge 2$; and $\lambda_0=\lambda_1=2$.
In all cases, the integral decays as a power of $N$, and since we have proved above
that $G_{01}$ also decays, we conclude that $G_{00} \to 1-V'(0)$
(with $V'(0)=0$ on the second-order curve and at the tricritical point by definition).

For the dense phase, we have $\partial_2F_{01} (t,t)= (1-V'(t))^2$ and
$\partial_2 F_{00}(t,t)= V''(t)$, and the result follows immediately from
Corollary~\ref{cor:Laplace-interior}.

By definition, on the first-order curve $V$ has global minima $V(0)=V(t_0)=0$,
with $t_0\neq 0$.
The hypotheses of Corollary~\ref{cor:Laplace-2min} hold
for $G_{01}$, with $\mu=1$ by the assumption that $V'(0)>0$ on
the first-order curve, and with $\lambda_0=1$ and $\lambda_1=2$
by \refeq{F01tt}.  The desired asymptotic formula for $G_{01}$ then follows
from \refeq{2min}.
Similarly, for the integral in \refeq{G00sum},
we have $\mu=1$, $\lambda_0=1$, $\lambda_1 \ge 2$,
so by Corollary~\ref{cor:Laplace-2min} the integral is asymptotic to a multiple
of $N^{-1/2}$.  It is therefore the constant term $1-V'(0)$
 in \refeq{G00sum} that dominates for $G_{00}$.
\end{proof}

\begin{proof}[Proof of Theorem~\ref{thm:mr}: susceptibility.]
It follows from Proposition~\ref{prop:2ptfcn} and
$\chi = G_{00} + (N-1)G_{01}$ that
the susceptibility obeys
\begin{align}
\lbeq{chiasy-pf}
    \chi & \sim
    \begin{cases}
    \frac{1-V'(0)}{V'(0)}\Gamma(2/1) & \text{(dilute phase)}
    \\
    N^{1/2}\frac{1}{(\frac{1}{2!} V''(0))^{1/2}} \Gamma(3/2) & \text{(second-order curve)}
    \\
    N^{2/3}\frac{1}{(\frac {1}{3!} V'''(0))^{1/3}}
    \Gamma(4/3) & \text{(tricritical point)}
    \\
    e^{-NV(t_0)}
    N^{1/2}\frac{\sqrt{2\pi}}{V''(t_0)^{1/2}}   & \text{(dense phase and first-order curve),}
    \end{cases}
\end{align}
as stated in Theorem~\ref{thm:mr}.
\end{proof}

We remark that
there is a mismatch for $G_{01}$ and for
the susceptibility as the dense phase approaches the second-order
curve.  For the susceptibility,
the limiting value from the dense phase (as $t_0 \downarrow 0$)
is $\chi \to N^{1/2}\sqrt{2\pi} (V''(0))^{-1/2}$, which is twice as big
as the value $N^{1/2}\Gamma(3/2) (\frac 12 V''(0))^{-1/2}
= N^{1/2}\frac 12 \sqrt{\pi} (\frac 12 V''(0))^{-1/2}$
on the second-order curve. The reason for this is clear from the proof:
in the dense phase the susceptibility receives a contribution from both sides of
the minimum of $V$ at $t_0$, whereas on the second-order curve it only receives a
contribution from the right-hand side of the minimum at $0$
(cf.\ the two insets at the second-order curve in Figure~\ref{fig:nuc}).

\subsection{Expected length}

Given the asymptotic behaviour for $\chi$ in \refeq{chiasy-pf},
the asymptotic formulas for the expected length stated in Theorem~\ref{thm:mr}
will follow once we prove that
\begin{align}
\lbeq{chiEL-pf}
    \chi\Ex L & \sim
    \begin{cases}
    \frac{\Vdot'(0)}{(V'(0))^2}
    & \text{(dilute phase)}  \vspace{1mm}
    \\
    N
     \frac{\Vdot'(0)}{ V''(0)}
      & \text{(second-order curve)}  \vspace{1mm}
      \\
      N^{4/3} \frac{\Gamma(2/3)}{3}
       \frac{1}{(\frac{1}{3!}V'''(0))^{2/3}}
       & \text{(tricritical point)}  \vspace{1mm}
       \\
      N^{3/2} e^{-NV(t_0)} \frac{\sqrt{2\pi}}{V''(t_0)^{1/2}}\Vdot(t_0)
      &
      \text{(dense phase including first-order curve).}
    \end{cases}
\end{align}
The proof of \refeq{chiEL-pf} is based on the following lemma.
Recall that $Q'=1-V'$.

\begin{lemma}
\label{lem:Ksim}
In the dilute phase, on the second-order curve,
at the tricritical point, and on the first-order curve
(the latter for the minimum of $V$ at $t=0$),
the forms $K_{0xy}$ defined in \refeq{K0xydef} have parameters $q_0,r_0,\lambda_0,\lambda_1$
as in Corollary~\ref{cor:Laplace-endpoint} given by:
\begin{alignat}{4}
    &\text{For $K_{012}$:}  &&\quad q_0= Q'(0)^2\Vdot'(0) ,
    &&\quad r_0 = \frac 12 Q'(0)^2 \Vdot'(0) ,
    &&\quad \lambda_0 = \lambda_1=2.
    \\
    &\text{For $K_{001}$:}  &&\quad q_0= Q'(0)\Vdot'(0),
    &&\quad
    r_0 =  Q'(0)\Vdot'(0),
    &&\quad \lambda_0 =  \lambda_1=1.
    \\
    &\text{For $K_{011}$:}  &&\quad q_0= Q'(0)\Vdot'(0) ,
    && &&\quad \lambda_0 = 1,\;   \lambda_1=2.
    \\
\lbeq{K000}
    &\text{For $K_{000}$:}  &&\quad q_0= \Vdot'(0),
    && &&\quad \lambda_0 = 0,\; \lambda_1=1.
\end{alignat}
\end{lemma}

\begin{proof}
The desired results can be read off from the following.

By \refeq{K0xydef},
$K_{012}(\Xinew^2,|\zeta|^2) = Q'(\Xinew^2)^2 
|\zeta|^2 \Vdot(\Xinew^2)$, so
\begin{equation}
    K_{012}(t,t) \sim Q'(0)^2
    \Vdot'(0)t^2,
    \qquad
    \partial_1 K_{012}(t,t)  \sim Q'(0)^2
    \Vdot'(0) t.
\end{equation}
Similarly,
$K_{001}(\Xinew^2,|\zeta|^2)
     =
    \big( \Vpm'(\Xinew^2) 
    +\Vpm'(\Xinew^2)^2 
    |\zeta|^2
    - V''(\Xinew^2)|\zeta|^2 \big)  \dot{V}(\Xinew^2)$
obeys
\begin{equation}
    K_{001}(t,t) \sim  \Vpm'(0)\Vdot'(0)t ,
    \qquad
    \partial_1  K_{001}(t,t) \sim \Vpm'(0)\Vdot'(0).
\end{equation}
Next,
$K_{011}(\Xinew^2,|\zeta|^2)
     =
    |\zeta|^2 \Vpm'(\Xinew^2) 
    \big(\Vpm'(\Xinew^2)
    \dot{V}(\Xinew^2)  + \dot{V}'(\Xinew^2)
     \big)$
obeys
\begin{align}
    K_{011}(t,t) &\sim  \Vpm'(0)\Vdot'(0)t ,
    \\
    \partial_1  K_{011}(t,t) &\sim
    \Big(\Vpm''(0)\Vdot'(0) + \Vpm'(0)^2\Vdot'(0) + \Vpm'(0)\Vdot''(0) \Big)t
    .
\end{align}
For the last case,
\begin{align}
    K_{000}(\Xinew^2,|\zeta|^2)
    & =
    \big( \Vpm'(\Xinew^2)
    + \Vpm'(\Xinew^2)^2
    |\zeta|^2 - V''(\Xinew^2)|\zeta|^2
     \big)\dot{V}(\Xinew^2)
     \nnb & \qquad +( 1 + 2\Vpm'(\Xinew^2)
    |\zeta|^2 ) \dot{V}'(\Xinew^2)
    +
    \dot{V}''(\Xinew^2)|\zeta|^2
\end{align}
obeys
\begin{align}
    K_{000}(t,t) &\sim  \Vdot'(0),
    \qquad
    \partial_1  K_{000}(t,t) \sim \Vpm'(0)\Vdot'(0) + \Vdot''(0).
\end{align}
This completes the proof.
\end{proof}

\begin{proof}[Proof of Theorem~\ref{thm:mr}: expected length.]
It suffices to prove \refeq{chiEL-pf}.
By Proposition~\ref{prop:EL},
\begin{align}
\lbeq{chiELintegrals}
    \chi \Ex L
    &=
    (N-1)(N-2)\int_{\R^2} D\Xinew\, e^{-NV(\Xinew^2)} K_{012}
    \nnb & \quad + (N-1) \int_{\R^2}D\Xinew\, e^{-NV(\Xinew^2)}(K_{001}+2K_{011})
    + \int_{\R^2} D\Xinew\, e^{-NV(\Xinew^2)}K_{000}.
\end{align}
It will turn out that
all three terms on the right-hand side contribute in the dilute
phase, but in all other cases only the first term contributes to the leading behaviour.
The integrability of \refeq{chiELintegrals} follows from
the lower bound on $V$ of Proposition~\ref{prop:Vderivs0}(i).

Consider first the dilute phase.
We apply Lemma~\ref{lem:Ksim} and
Corollary~\ref{cor:Laplace-endpoint} with $\mu=1$, $v_0=V'(0)$ and immediately obtain
\begin{align}
\lbeq{K012dilute}
    \int_{\R^2} D\Xinew\, e^{-NV(\Xinew^2)}K_{012}
    & \sim  \frac{ (1-V'(0))^2 \Vdot'(0)}{(V'(0)N)^2} ,
    \\
\lbeq{K001dilute}
    \int_{\R^2} D\Xinew\, e^{-NV(\Xinew^2)}K_{001}
    &   = o(N^{-1}),
    \\
\lbeq{K011dilute}
    \int_{\R^2} D\Xinew\, e^{-NV(\Xinew^2)}K_{011}
    & \sim  \frac{ (1-V'(0)) \Vdot'(0)}{V'(0)N},
    \\
\lbeq{K000dilute}
    \int_{\R^2} D\Xinew\, e^{-NV(\Xinew^2)}K_{000}
    & \sim \Vdot'(0).
\end{align}
Therefore, in the dilute phase, as stated in \refeq{chiEL-pf},
\begin{equation}
    \chi \Ex L \sim
    \Vdot'(0) \left(
    \frac{(1-V'(0))^2}{(V'(0))^2}  + 2 \frac{1-V'(0)}{V'(0)} + 1
    \right)
    =
    \frac{\Vdot'(0)}{(V'(0))^2}.
\end{equation}

Consider next the second-order curve ($\mu=2$, $v_0=\frac{1}{2!}V''(0)$)
and the tricritical point ($\mu=3$, $v_0=\frac{1}{3!}V'''(0)$).
For these cases, Lemma~\ref{lem:Ksim} and
Corollary~\ref{cor:Laplace-endpoint} give
\begin{equation}
    \int_{\R^2} D\Xinew\, e^{-NV(\Xinew^2)}K_{012}(\Xinew^2,|\zeta|^2)
    \sim
    \frac{(1-V'(0))^2 \Vdot'(0)}{\mu(v_0N)^{2/\mu}}  \Gamma(2/\mu)
     .
\end{equation}
By definition, on the second-order curve $V'(0)=0$, and at the tricritical point $V'(0)=0$
and $\Vdot'(0)=M_1=1$ (recall Proposition~\ref{prop:Vderivs0}(ii)).
By Lemma~\ref{lem:Ksim} and Corollary~\ref{cor:Laplace-endpoint},
the integrals involving $K_{001}$ and $K_{011}$ are at most $O(N^{-1/\mu})$,
and the one involving $K_{000}$ is at most $O(1)$.
Since the latter are multiplied
by $N$ and $1$ respectively, these terms contribute order $N^{1-1/\mu}$ and $N^0$,
and this is less than the $K_{012}$ term
which is multiplied by $N^2$ and hence is order $N^{2-2/\mu}$.
This proves the second-order and tricritical cases of \refeq{chiEL-pf}.

Next, we consider the dense phase.
Let $t_0>0$ be the location of the global minimum of $V$.  We have $V(t_0) <0$,
$V'(t_0)=0$,  and $V''(t_0)>0$.
By Corollary~\ref{cor:Laplace-interior}
(note that $V$ and the various $K_*$ satisfy the analyticity hypotheses by definition),
\begin{align}
    \int_{\R^2} D\Xinew\, e^{-NV(\Xinew^2)}K_{0xy}
    & \sim
    e^{-NV(t_0)}
    \frac{A_{0xy}}{N^{1/2}}
    ,
    \qquad
    A_{0xy}  = \frac{\sqrt{2\pi}}{V''(t_0)^{1/2}}
    \partial_2 K_{0xy}(t_0,t_0)   .
\end{align}
There are order $N^2$ terms with $0,x,y$ distinct, order $N$ terms where only two are
distinct, and a single term where $0=x=y$.  Since each term has the same $N^{-1/2}e^{-NV(t_0)}$
behaviour,
only the case with $0,x,y$ distinct can contribute to $\chi\Ex L$.
Since $K_{012}=(1-V'(\Xinew^2))^2\Vdot(\Xinew^2)|\zeta|^2$ obeys
\begin{align}
    \partial_2K_{012}(t_0,t_0) &= (1-V'(t_0))^2\Vdot(t_0) = \Vdot(t_0),
\end{align}
Corollary~\ref{cor:Laplace-interior} gives
\begin{align}
    \chi\Ex L & \sim
     N^2
    e^{-NV(t_0)}
    \frac{ 1}{N^{1/2}}\frac{\sqrt{2\pi}}{V''(t_0)^{1/2}}
    \Vdot(t_0),
\end{align}
as stated in \refeq{chiEL-pf}.

Finally, we consider the first-order curve, with global minima $V(0)=V(t_0)=0$.
We apply Corollary~\ref{cor:Laplace-2min} to each of the integrals in
\refeq{chiELintegrals}, using $\mu=1$ and the values of $\lambda_i$ stated in
Lemma~\ref{lem:Ksim}.  After taking into account the $N$-dependent factors in
the terms of \refeq{chiELintegrals}, we conclude from Corollary~\ref{cor:Laplace-2min}
that $\chi\Ex L$ has the same asymptotic behaviour on the first-order curve as it
does in the dense phase, now with $V(t_0)=0$, namely
\begin{align}
    \chi\Ex L & \sim
     N^{3/2}\frac{\sqrt{2\pi}}{V''(t_0)^{1/2}}
    \Vdot(t_0).
\end{align}
This completes the proof.
\end{proof}

\section{Phase diagram for the example: proof of Theorem~\ref{thm:exponents}}
\label{sec:phase-diagram-pfs}

In this section we prove Theorem~\ref{thm:exponents}, which concerns the
differentiability of the phase boundary $\nu_c(g)$ at the tricritical point $g_c$,
and the asymptotic behaviour of the susceptibility and density,
for the specific example $p(t) = e^{-t^3-gt^2-\nu t}$.
According to \refeq{vintegral},
the effective potential is the function $V:[0,\infty)\to \R$
defined by
\begin{equation}
\lbeq{Vp}
    V(t) = t - \log (1+v(t)), \qquad
    v(t) = \int_0^\infty e^{-t^3-gt^2-(\nu+1) t} \sqrt{\frac{t}{s}}I_1(2\sqrt{ts}) ds.
\end{equation}
We emphasise that $V$ is a function of a single real variable, so
in principle complete information could be extracted with sufficient effort.

\subsection{Numerical input}
\label{sec:numericalinput}

\begin{figure}[ht]
\centering{
\includegraphics[scale=1.0]{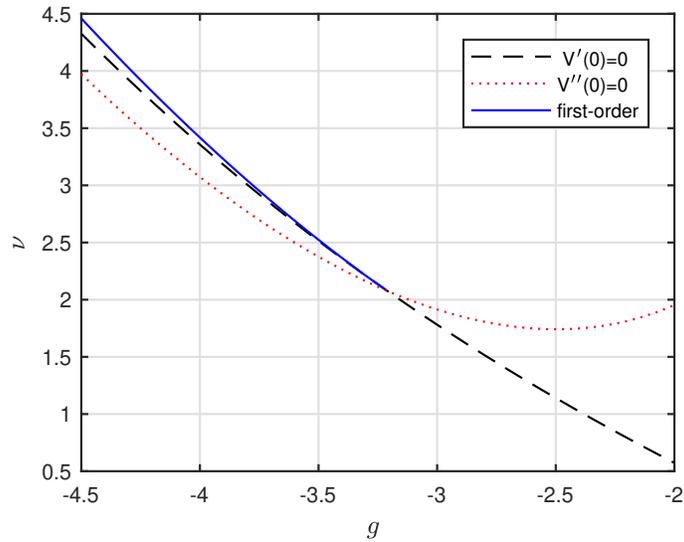}
\caption{Phase diagram for $p(t) = e^{-t^3-gt^2-\nu t}$.}
\label{fig:nuc-bis}
}
\end{figure}

As discussed before its statement,
our proof of Theorem~\ref{thm:exponents} relies on a numerical analysis of
the effective potential \refeq{Vp}, whose conclusions are summarised
in Figure~\ref{fig:nuc} which for convenience we repeat here
in abridged form as Figure~\ref{fig:nuc-bis}.

The dashed (black) and dotted (red) curves in Figure~\ref{fig:nuc-bis} are respectively the
curves defined implicitly by
$V'(0)=1-M_0=0$ and $V''(0)=M_0^2-M_1=0$.  Above the dashed curve $V'(0)>0$ and
below the dashed curve $V'(0)<0$.  Above the dotted curve $V''(0)<0$ and below the
curve $V''(0)>0$.
Below the curve $V'(0)=0$ there is a unique solution to $V'(t_0)=0$.
The two curves $V'(0)=0$ and $V''(0)=0$ intersect at the tricritical point, which is
\begin{equation}
    g_c = -3.2103..., \quad \nu_c = 2.0772... \, .
\end{equation}
The solid curve (for $g \le g_c$) is the first-order curve and the dashed curve
(for $g \ge g_c$) is the second-order curve.  These two curves satisfy the
conditions of Definition~\ref{defn:phases}.
Together the first- and second-order curves define the phase boundary $\nu_c(g)$.
Points below
the phase boundary are in the dense phase
in the sense of Definition~\ref{defn:phases}, and points above
the phase boundary are in the dilute phase in the sense of Definition~\ref{defn:phases}.
The first few moments at the tricritical point are
\begin{align}
\lbeq{Mnumbers}
    &M^c_0=M^c_1=1, \quad M^c_2 = 1.4478..., \quad M^c_3 = 2.4062..., \quad M^c_4 = 4.3315....
\end{align}
To distinguish moments at the tricritical point $(g_c,\nu_c)$ from moments computed at
other points $(g,\nu)$, we write the former as $M_i^c$ and the latter simply as $M_i$.

\subsection{First derivative of phase boundary}

We prove Theorem~\ref{thm:exponents}(i) in two lemmas.
Together, the lemmas show that the phase boundary is
differentiable but not twice differentiable at the
tricritical point with derivative $-M_2^c$.
In this section, we prove the differentiability.  The proof of the inequality of the
left and right second derivatives at the tricritical point is deferred to
Lemma~\ref{lem:boundary-quadratic}.
Here and throughout Section~\ref{sec:phase-diagram-pfs}, we use the following elementary facts
about derivatives of moments, for $i \ge 0$:
\begin{align}
\lbeq{Mderivs}
    M_{i,g}=-M_{i+2}, \quad M_{i,\nu}=-M_{i+1}, \quad M_{i,gg}= M_{i+4},
    \quad M_{i,g\nu}=M_{i+3}, \quad M_{i,\nu\nu}=M_{i+2}.
\end{align}

\begin{lemma}
\label{lem:tangent}
(i) The tangent to the curve $M_0=1$ (the entire dashed curve in Figure~\ref{fig:nuc-bis},
including  the second-order curve and the tricritical point) has slope $\nu_g =-M_2/M_1$.
\\
(ii)
The tangent to the first-order curve at the tricritical point also has slope
$\nu_{c,g}=-M^c_2/M_1^c=-M_2^c$.
\end{lemma}

\begin{proof}
(i) We write the curve $M_0=1$ as $\nu=\nu(g)$.
Implicit differentiation of
$M_0(g,\nu(g))=1$
with respect to $g$, together with \refeq{Mderivs},
gives $0=M_{0,g} + M_{0,\nu} \nu_g = -M_2 -M_1\nu_g$,
so $\nu_g = -M_2/M_1$.

\smallskip\noindent(ii)
On the first-order curve $\nu=\nu_c(g)$ (for $g\le g_c$),
by Definition~\ref{defn:phases} there are two solutions to $V(t_0)=0$:
$t_0=0$ and a positive $t_0>0$, which by definition characterises the first-order curve.
At the positive root, $V'(t_0)=0$.
At the tricritical point, $t_0=0$ and $V(0)=V'(0)=V''(0)=0$.
By continuity, as $g\uparrow g_c$ we have $t_0 \to 0$.
Total derivatives with respect to $g$ are denoted $\mathring{f} = \frac{d}{dg}f$.

We differentiate $V(t_0(g,\nu_c(g)),g,\nu_c(g))=0$ with respect to $g$
and obtain
\begin{align}
\lbeq{Vring}
    V'\mathring{t}_0 + V_g + V_\nu \nu_{c,g} & = 0 .
\end{align}
For every $g\le g_c$,
$V'(t_0(g,\nu_c(g)),g,\nu_c(g))=0$, so the first term is constant in $g$ and is equal to zero.
Also, $V_\nu$ is nonzero on the first-order curve, so $\nu_{c,g} = -V_g/V_\nu$.
We parametrise the first order curve as $(g(s),\nu(s))=(g_c-s,\nu_c(g_c-s))$ for
$s\geq 0$.  By definition, the slope of the tangent to the first-order curve is $\lim_{s\downarrow 0} \nu_{c,g}$.
Also, by definition, by \refeq{v-expansion}, and by \refeq{Mderivs}, as $s\downarrow 0$ we have
\begin{equation}
    \frac{V_g}{V_\nu} = \frac{v_g}{v_\nu} \sim \frac{M_{0,g}t_0}{M_{0,\nu}t_0}
    =
    \frac{M_2}{M_1} \sim M_2^c,
\end{equation}
where we also used $M_1^c=1$ and
the fact that $t_0(g(s),\nu(s)) \to 0$ as $s\downarrow 0$.
Therefore, at the tricritical point, $\nu_{c,g} =-M_2^c$.
\end{proof}

\subsection{Susceptibility}

We now prove Theorem~\ref{thm:exponents}(ii).
Given a point $(g(0),\nu(0))$ on the second-order curve, or the tricritical point,
we fix a vector ${\bf m}=(m_1,m_2)$ with base at $(g(0),\nu(0))$,
which is nontangential to the second-order curve and
pointing into the dilute phase.  By Lemma~\ref{lem:tangent},
${\bf n}=(M_2,M_1)$ is normal to the curve and pointing into the dilute phase, so
${\bf m}\cdot {\bf n}>0$
(moments are evaluated at $(g(0),\nu(0))$).
We define a line segment in the dilute phase that starts at our fixed point by
\begin{equation}
\lbeq{segment}
    (g(s),\nu(s)) = (g(0)+sm_1,\nu(0)+sm_2)
    \qquad (s\in [0,1]),
\end{equation}
and set $\chi(s)=\chi(g(s),\nu(s))$.
We set  $M_0(s)=M_0(g(s),\nu(s))$ and define other functions similarly.

By \refeq{chiasy-mr}, the infinite-volume susceptibility in the dilute phase is given by
\begin{equation}
    \chi = \frac{1-V'(0)}{V'(0)} = \frac{M_0(s)}{1-M_0(s)}.
\end{equation}
By \refeq{Mderivs} and the chain rule, $M_0(s) \sim 1-(M_2(0)m_1+M_1(0)m_2)s
=1-({\bf m}\cdot {\bf n})s$ as $s \downarrow 0$.
This gives
\begin{equation}
    \chi \sim
    \frac{1}{({\bf m}\cdot {\bf n})s},
\end{equation}
which proves that $\chi$ diverges as stated in \eqref{e:chiasy-thm}.

\subsection{Density}

The effective potential $V$ is smooth in $(g,\nu)$, has a uniquely attained global
minimum at $t=0$ on the second-order curve (with $V(0)=0$), has a uniquely
attained global minimum at $t_0>0$ in the dense phase (with $V(t_0)<0$),
and attains its global minimum at both $0$ and $t_0$ on the first-order
curve (with $V(0)=V(t_0)=0$).
By smoothness of $V$, $t_0 \to 0$ as the second-order curve or tricritical point is approached.
In the dense phase, the density is given by $\rho = \Vdot(t_0)$,
so as $t_0\to 0$,
$\Vdot(t_0) \sim \Vdot'(0)t_0  = M_1 t_0$, and at the tricritical point $M_1^c=1$.

To prove Theorem~\ref{thm:exponents}(iii), it therefore suffices to prove
the following Propositions~\ref{prop:t0i}, \ref{prop:t0ii} and \ref{prop:t0iii}
for the asymptotic behaviour of $t_0$.
The values of the constants $A,B_i$ in the propositions are specified in their proofs.
The fact that the phase boundary is not twice differentiable at the tricritical
point is proved in Lemma~\ref{lem:boundary-quadratic}, to complete the proof
of Theorem~\ref{thm:exponents}(i).

Let $(g(0),\nu(0))$ be a given point on the second-order curve, or the tricritical point.
We again use the normal ${\bf n}=(M_2,M_1)$ which points into the dilute phase.
As in \refeq{segment} we fix a vector ${\bf m} = (m_1,m_2)$,
but now with ${\bf m}\cdot{\bf n}<0$ so that ${\bf m}$
points into the dense phase, and we define a line segment that starts at our given point by
\begin{equation}
\lbeq{segment2}
    (g(s),\nu(s)) = (g(0)+sm_1,\nu(0)+sm_2)
    \qquad (s\in [0,1]).
\end{equation}
We consider the asymptotic behaviour of the density $\rho$ along this segment,
as $s \downarrow 0$.
Let
$M_i(s) = M_i(g(s),\nu(s))$ for $i =0,1,2,\ldots$.
An ingredient in the proofs is the asymptotic formula, as $s \downarrow 0$,
\begin{equation}
\lbeq{MTay}
    M_i(s) = M_i - (m_1M_{i+2} +m_2M_{i+1})s +
    \frac 12
    (m_1^2 M_{i+4} + 2m_1m_2 M_{i+3}+m_2^2 M_{i+2})s^2 + O(s^3),
\end{equation}
with all moments on the right-hand side evaluated at $s=0$.
This follows
from Taylor's theorem and \refeq{Mderivs}.

\subsubsection{Approach to second-order curve from dense phase}

\begin{prop}
\label{prop:t0i}
As the second-order curve is approached,
\begin{equation}
  t_0 \sim
  \begin{cases}
    \frac{|{\bf m} \cdot {\bf n}|}{1-M_1}s
    & \text{(nontangentially from dense phase)}
  \\
  A s^2 & \text{(tangentially from dense phase)}.
  \end{cases}
\end{equation}
The constant $A$ is strictly positive at least
along some arc of the second-order curve adjacent to the tricritical point.
\end{prop}

\begin{proof}
We parametrise the approach to a point $(g(0),\nu(0))$ on the second order curve
as $(g(s),\nu(s))$ with $s \downarrow 0$, as in \refeq{segment2},
and we write $V_s(t) = V(g(s),\nu(s); t)$.
On the second order curve, $V'_0(0)=0$, $V''_0(0)>0$ and $0$ is a global minimum
of $V_0$. On the other hand, for $s>0$
there is a unique solution $t_0=t_0(s)>0$ to $V'_s(t_0(s))=0$.
The condition $V'(t_0)=0$ is equivalent to $1+v(t_0) = v'(t_0)$, so
\begin{equation}
  1+v(0)+v'(0)t_0 = v'(0)+v''(0)t_0 + O(t_0^2).
\end{equation}
By \refeq{v-expansion}, this gives
\begin{equation}
    1+ M_0t_0 = M_0 + M_1 t_0  + O(t_0^2) .
\end{equation}
  Therefore, since $t_0=o_s(1)$, $1-M_1(0) >0$, and $M_0(s)-1 = o_s(1)$,
\begin{equation}
\lbeq{t0i}
  t_0 = \frac{M_0-1}{M_0-M_1}(1+O(t_0)) \sim \frac{M_0-1}{1-M_1(0)}
  .
\end{equation}

\smallskip\noindent
\emph{Second-order curve nontangentially from dense phase.}
By \refeq{t0i} and \refeq{MTay},
\begin{equation}
  t_0   \sim \frac{|{\bf m}\cdot {\bf n}|}{1-M_1(0)}s.
\end{equation}
This proves the result for the nontangential approach, for which ${\bf m}\cdot {\bf n}< 0$.

\smallskip\noindent
\emph{Second-order curve tangentially from dense phase.}
For the tangential approach, we choose ${\bf m} = (M_1,-M_2)$ so that
${\bf m}\cdot {\bf n}=0$.
By \refeq{MTay} we have
$M_0(s) \sim 1 + a s^2$ with
$a=\frac 12 (M_1^2M_4-2M_1M_2M_3+M_2^3)$, so now
\begin{equation}
\lbeq{t02tan}
t_0 \sim
\frac{a}{1-M_1(0)}s^2
\end{equation}
(all moments are evaluated at $s=0$ here).
At the tricritical point, $a^c= M_4^c-2M_2^c M_3^c + (M_2^c)^3>0$ by
\refeq{Mnumbers} (cf.\ Lemma~\ref{lem:boundary-quadratic}).
By continuity, $a$
remains positive at least along some arc of the
second-order curve adjacent to the tricritical point.
The constant $A$ is $A=(1-M_1)^{-1}a$.
\end{proof}

\subsubsection{Approach to tricritical point from dense phase}

\begin{figure}[ht]
\centering{
\includegraphics[scale=0.7]{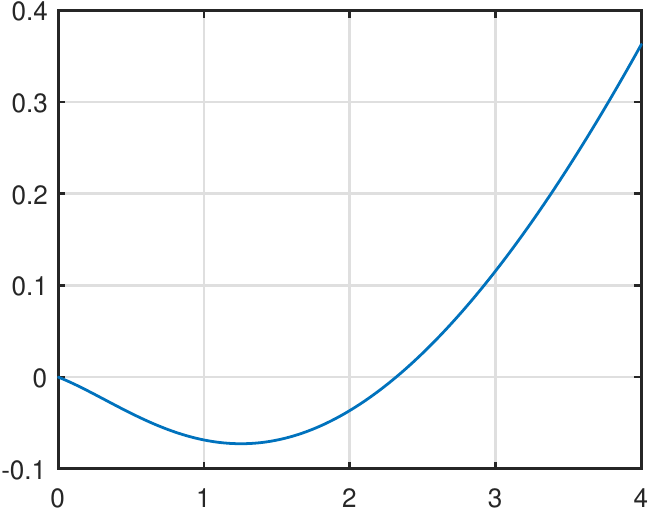}

\caption{Effective potential $V$ vs $t$ on tangent line (first-order side), at $(g,\nu)=(-3.700,2.786)$.
The point $t_0$ is the location of the minimum.}
\label{fig:Veg2}}
\end{figure}

We parametrise the approach to the tricritical point $(g_c,\nu_c)$
as $(g(s),\nu(s))$ with $s \downarrow 0$, as in \refeq{segment2},
and we write $V_s(t) = V(g(s),\nu(s); t)$.
For both tangential and nontangential
approach to the tricritical point from the dense phase, the
approach is from below the curve $V'(0)=1-M_0=0$ and there is one solution $t_0=t_0(s)>0$ to $V_s'(t_0)=0$. An example is depicted in Figure~\ref{fig:Veg2}.
For this approach, $V_s'(0) =1-M_0<0$, and $V_s''(0)=M_0^2-M_1$ can have either sign or equal zero
(the dotted curve in Figure~\ref{fig:nuc-bis} is the curve $V''(0)=0$).

\begin{prop}
\label{prop:t0ii}
As the tricritical point is approached,
\begin{equation}
    t_0 \sim
    \begin{cases}
    B_0
    ( |{\bf m} \cdot {\bf n}|s)^{1/2} & \text{(nontangentially from dense phase)}
    \\
    B_1
    s & \text{(tangentially from second-order side)}
    \\
    B_2
    s & \text{(tangentially from first-order side})
    .
    \end{cases}
\end{equation}
The constants
$B_0,B_1,B_2$ are all strictly positive.
\end{prop}

The proof of Proposition~\ref{prop:t0ii} uses the following elementary lemma,
which gives an estimate for how fast $t_0 \to 0$.

\begin{lemma}
  \label{lem:Newtonnew}
For $s\in [0,1]$, let $f_s:[0,\infty) \to \R$ be smooth functions, with $f_s$ and
its derivatives uniformly continuous (and hence uniformly bounded)  in small $s,t$.
Suppose that $f_0(0)=f_0'(0)=0$, $f_0''(0)>0$, and that $f_s(0)< 0$ for $s>0$.
Then for small $s>0$ there is a unique root $t_0=t_0(s)$ of $f_s(t_0(s))=0$ and
as $s \downarrow 0$,
\begin{equation}
  t_0(s) = O(|f_s(0)|^{1/2}+|f_s'(0)|).
\end{equation}
\end{lemma}

\begin{proof}
  By the assumptions that $f_0''(0)>0$ and that $f_s''(t)$ is uniformly continuous
  in small $s,t$, there are $\delta,c>0$
  such that $f_s''(t)\geq 2c$ for $s,t\leq \delta$, and hence
  $f_s(t) \geq f_s(0) +f_s'(0)t+ct^2 = -|f_s(0)|+f_s'(0)t+ct^2$ for $s,t \leq \delta$.
  Since the positive roots to $-a+bt+ct^2 =0$ with $a,c>0$ are
  \begin{equation}
    t = \frac{1}{2c} (-b+ \sqrt{b^2+4ac})
    \leq \frac{1}{2c} (2\max\{-b,0\} + 2\sqrt{ac})
    =
    \frac{\max\{-b,0\}}{c} + \sqrt{\frac{a}{c}},
  \end{equation}
  we see that $f_s(t) >0$ if $t > \sqrt{|f_s(0)|/c} + \max\{-f_s'(0),0\}/c$.
  In particular, for $s$ small enough that $\sqrt{|f_s(0)|/c}+|f_s'(0)|/c \leq \delta$,
  it follows by continuity and $f_s(0)<0$ that $f_s$ has a root $t_0=t_0(s) \in [0,\delta]$ to $f_s(t_0)=0$
  satisfying $t_0 = O(\sqrt{|f_s(0)|}+|f'_s(0)|)$. Since $f_s$ is convex on $[0,\delta]$, this root is unique.
\end{proof}

\begin{proof}[Proof of Proposition~\ref{prop:t0ii}]
We apply Lemma~\ref{lem:Newtonnew} to $f_s(t)=V'_s(t)$.
The hypotheses are satisfied:  $f_s(0)=1-M_0(s)<0$, $f_0'(0)=0$, and by \eqref{e:Vderivs0}
and \refeq{Mnumbers},
\begin{equation} \label{e:alpha}
    \alpha = f_0''(0)=V'''_0(0) = -\frac 12 M_2^c  + 1 >0.
\end{equation}
Taylor expansion of $V'(t_0)=0$ gives
\begin{equation}
        V'(0) + V''(0)t_0 + \frac 12 V'''(0) t_0^2 +O(t_0^3) =0.
\end{equation}
It follows from Lemma~\ref{lem:Newtonnew} that
\begin{align}
\lbeq{t0ii}
    t_0 & =
    \frac{-V''(0) \pm \sqrt{V''(0)^2 - 2 V'''(0) (V'(0)+O(|V'(0)|^{3/2}+|V''(0)|^3)}}{V'''(0)}
\end{align}

\smallskip\noindent
\emph{Tricritical point nontangentially from dense phase.}
By \refeq{MTay},  in this case we have
\begin{equation}
    V_s'(0)=1-M_0(s) \sim -|{\bf m} \cdot {\bf n}|s,
    \qquad
    V_s''(0) \sim 1-M_1(s) =O(s).
\end{equation}
Therefore the error term in \refeq{t0ii} is $O(s^{3/2})$ and hence
\begin{align}
  t_0
  & \sim
    \alpha^{-1}(2\alpha|{\bf m} \cdot {\bf n}|s)^{1/2}.
\end{align}
The constant $B_0$ is $B_0=(2\alpha^{-1})^{1/2} = 2 (2-M_2^c)^{-1/2}$.

\smallskip
\noindent\emph{Tricritical point tangentially from second-order side.}
The slope of the tangent line at the tricritical point is $-M_2^c$, so
the tangential approach is parametrised
by \refeq{segment2} with ${\bf m} = (1,-M_2^c)$.
On this tangent line, $M_0>1$ and  $M_0^2-M_1 >0$
(see Figure~\ref{fig:nuc-bis}).
As above \refeq{t02tan},
$M_0(s) \sim 1+a^cs^2$ with
$a^c=\frac 12 ((M_1^c)^2M_4^c -2M_1^c M_2^c M_3^c +(M_2^c)^3)>0$, and
by \refeq{MTay} $M_1(s) \sim 1-bs$ with $b=M_3^c-(M_2^c)^2>0$ by \refeq{Mnumbers}.
Therefore,
\begin{equation}
\lbeq{Vab}
    V_s'(0) = 1-M_0(s) \sim -a^cs^2,
    \qquad
    V_s''(0) = M_0(s)^2-M_1(s) \sim bs.
\end{equation}
As in the previous case, we apply Lemma~\ref{lem:Newtonnew}
and \refeq{t0ii} but this time with
an error term which is $O(s^3)$.  This leads to
\begin{equation}
    t_0 \sim \alpha^{-1}\Big(- bs + \sqrt{b^2s^2 + 2\alpha a s^2} \Big)
    =
    \alpha^{-1}\Big(- b + \sqrt{b^2 + 2\alpha a } \Big)s.
\end{equation}
The constant $B_1$ is
$B_1=\alpha^{-1}( -b + \sqrt{b^2 + 2\alpha a }) >0$.

\smallskip
\noindent
\emph{Tricritical point tangentially from first-order side.}
Now we parametrise the tangential approach
as in \refeq{segment2} with ${\bf m} = (-1,M_2^c)$.
Now $M_0>1$ and $M_0^2-M_1<0$ (see Figure~\ref{fig:nuc-bis}).
The formula \refeq{t0ii} again applies with error term $O(s^3)$.
Again $M_0(s) \sim 1+as^2$, but now
\begin{equation}
\lbeq{Mb}
    M_1(s)\sim 1+bs
\end{equation}
due to the
replacement of ${\bf m}$ by $-{\bf m}$.
Therefore,
\begin{equation}
    t_0 \sim \alpha^{-1}\Big( b
    s + \sqrt{b^2s^2 + 2 \alpha a s^2} \Big)
    =
    \alpha^{-1}\Big( b
    + \sqrt{b^2 + 2\alpha a } \Big)s.
\end{equation}
The constant $B_2$ is
$B_2 = \alpha^{-1}(b+ \sqrt{b^2 + 2\alpha a } )>0$.
\end{proof}

\subsubsection{Approach to tricritical point along first-order curve}
\label{sec:1storder}

\begin{figure}[ht]
\centering{
\includegraphics[scale=0.7]{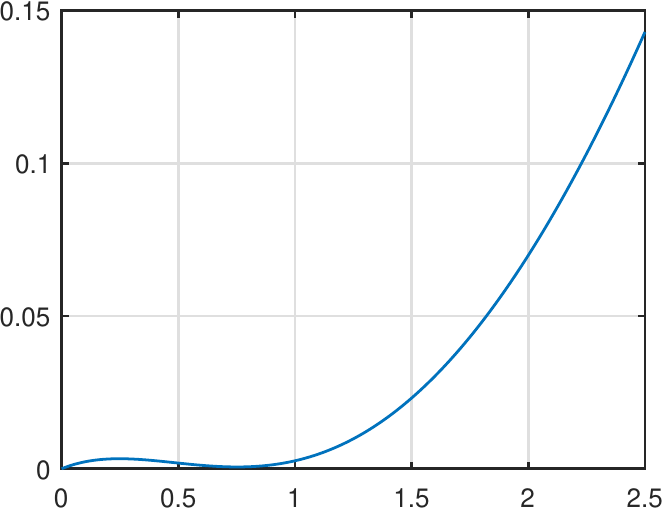}

\caption{Effective potential $V$ vs $t$ on
first-order curve, at $(g,\nu)=(-3.700,2.864)$.
The point $t_0$ is the unique positive solution to $V(t_0)=0$.
}
\label{fig:Veg1}}
\end{figure}

We now complete the proof of Theorem~\ref{thm:exponents}, by proving
the last case of Theorem~\ref{thm:exponents}(iii) in Proposition~\ref{prop:t0iii} and
the remaining part of Theorem~\ref{thm:exponents}(i) (namely the
failure of the phase boundary to be twice differentiable)
in Lemma~\ref{lem:boundary-quadratic}.
The focus is on the location $t_0$ of the nonzero
minimum of the effective potential on the first-order curve.
Figure~\ref{fig:Veg1} shows a typical $V$.

\begin{prop}
\label{prop:t0iii}
As the tricritical point is approached along the
first-order curve $(g(s),\nu(s))=(g_c-s,\nu_c(g_c-s))$,
\begin{equation}
\lbeq{t0B3}
    t_0 \sim
    B_3
    s \qquad \text{(along first-order curve})
    .
\end{equation}
The constant
$B_3$ is strictly positive.
\end{prop}

\begin{proof}
On the first-order curve, $V_s'(0)>0$ and  $V_s''(0)<0$ (see Figure~\ref{fig:nuc-bis}),
and there is a unique $t_0>0$ such
that $V_s(t_0)=0$. At this minimum, $V_s'(t_0)=0$ and $V_s''(t_0) >0$.
Under Taylor expansion, together with the fact that
$V_s(0)=0$, the equations $V_s(t_0)=0$ and $V_s'(t_0)=0$ become
\begin{align}
\lbeq{Vtay}
    0 &=
    t_0
        \left( V_s'(0) + \frac{1}{2!} V_s''(0)t_0 + \frac{1}{3!} V_s'''(0)t_0^2 +O(t_0^3) \right)
  \\
\lbeq{Vptay}
      0 & =
          V_s'(0) + V_s''(0)t_0 + \frac{1}{2!} V_s''(0)t_0^2 +O(t_0^3)
        .
\end{align}

By the chain rule,
$\frac{d}{ds}\nu_c(g(s)) = -\nu_{c,g}(g(s))$, and
by Lemma~\ref{lem:tangent},
$-\nu_{c,g}(g(s)) \to M_2^c$ as $s \downarrow 0$.
Therefore, as in \refeq{Mb},
\begin{equation}
\lbeq{Mb1}
    M_1(s) \sim 1+bs,
\end{equation}
with $b=M_3^c-(M_2^c)^2>0$.
Also, since
\begin{align}
    \frac{d}{ds}M_0(s) & =  -M_{0,g}(s)-M_{0,\nu}(s) \nu_c'(g_c-s) \to
    M_2^c -M_1^c(-M_2^c/M_1^c) = 0
    \qquad (s \to 0),
\lbeq{Mpp}
\end{align}
it follows that
\begin{equation}
\lbeq{Vab1}
    V_s''(0) = M_0(s)^2-M_1(s) \sim -bs.
\end{equation}
We substitute this into \refeq{Vtay}--\refeq{Vptay} and use
$V'''(0)=\alpha =1 - \frac 12 M_2^c >0$, to find that
\begin{align}
  V'(0)-\frac{bs}{2}t_0+\frac{\alpha}{6}t_0^2 = O(t_0(t_0+s)^2),
  \\
  V'(0)- bs t_0+\frac{\alpha}{2}t_0^2 = O(t_0(t_0+s)^2),
\end{align}
and hence
\begin{equation} \label{e:kappatringasymp}
  V'(0)- \frac{bs}{4}t_0 =  O(t_0(t_0+s)^2).
\end{equation}
We substitute \eqref{e:kappatringasymp}
into either of \refeq{Vtay}--\refeq{Vptay}, and
after cancellation of a factor $t_0$, we obtain
\begin{equation}
  -\frac{bs}{4}+\frac{\alpha}{6}t_0 = O(t_0+s)^2.
\end{equation}
Since $t_0 = o_s(1)$, this gives
\begin{equation} \label{e:t0asymp}
  t_0 \sim \frac{3b}{2\alpha} s = B_3 s,
\end{equation}
so $B_3 =\frac{3b}{2\alpha}=3\frac{M_3^c - (M_2^c)^2}{2-M_2^c} >0$ and the proof is complete.
\end{proof}

Note that the conclusion \refeq{t0B3} agrees with the naive argument
(ignoring the error term) that
the quadratic in \refeq{Vtay} will have a unique root precisely when the discriminant vanishes,
i.e., when
\begin{equation}
    V_s'(0) = \frac 38 \frac{V_s''(0)^2}{V_s'''(0)},
\end{equation}
and in this case the root is
$t_0 = \frac{3}{2}\frac{|V_s''(0)|}{V_s'''(0)} \sim \frac 32 \frac{bs}{\alpha}$.

Finally, we prove the remaining part of Theorem~\ref{thm:exponents}(iii),
concerning the behaviour of the density as the tricritical point is approached
along the first-order curve.

\begin{lemma}
\label{lem:boundary-quadratic}
(i)
At the tricritical point,
the second derivative of the second-order curve
is given by $\nu_{gg} = M_4^c -2M_3^c M_2^c +(M_2^c)^3 > 0$.
\\ (ii)
At the tricritical point, the second derivative of the first-order curve is
instead $\nu_{c,gg}=M_4^c - 2M_3^cM_2^c +   (M_2^c)^3 +
3b^2/(4\alpha)$,
with $b = M_3^c-(M_2^c)^2 >0$ and with $\alpha = 1- \frac 12 M_2^c>0$.
\end{lemma}

\begin{proof}
(i)
The second-order curve $\nu=\nu(g)$ is given by $V'(0)=0$.  Since $V'(0)=1-M_0$,
a first differentiation
with respect to $g$ gives $0=M_{0,g}+M_{0,\nu}\nu_g$, and a second differentiation gives
\begin{align}
    0 & = M_{0,gg} + 2M_{0,\nu g}\nu_g   + M_{0,\nu\nu} \nu_g^2 + M_{0,\nu} \nu_{gg}.
\end{align}
We use $\nu_g = -M_2/M_1$ from Lemma~\ref{lem:tangent}(i), \refeq{Mderivs}, and $M_1^c=1$ to see that,
at the tricritical point,
\begin{equation}
\lbeq{nugg}
    \nu_{c,gg}  = M_4^c - 2M_3^cM_2^c +   (M_2^c)^3.
\end{equation}
By \refeq{Mnumbers}, $\nu_{c,gg} >0$.

\smallskip\noindent (ii)
By \refeq{Vring}, on the first-order curve we have $V_g+V_\nu\nu_{c,g}=0$.
Total derivatives with respect to $g$ are denoted $\mathring{f} = \frac{d}{dg}f$.
Differentiation with respect to $g$ gives
\begin{align}
    \mathring{V}_g + \mathring{V}_\nu \nu_{c,g} + V_\nu \nu_{c,gg} & = 0,
\end{align}
so, since $V_\nu \sim M_1^ct_0= t_0$ by \refeq{v-expansion},
\begin{equation}
    \nu_{c,gg}  = - \frac{1}{V_\nu}(\mathring{V}_g + \mathring{V}_\nu \nu_{c,g})
      \sim - \frac{1}{t_0}(\mathring{V}_g + \mathring{V}_\nu \nu_{c,g})
    .
\end{equation}
We need to compute the coefficient of $t_0$ in
$\mathring{V}_g + \mathring{V}_\nu \nu_{c,g}$.
Since
\begin{align}
    \mathring{V}_g & = V_g'\mathring{t}_0 + V_{gg} + V_{g\nu}\nu_{c,g}
    ,
    \\
    \mathring{V}_\nu & = V_\nu'\mathring{t}_0 + V_{\nu g} + V_{\nu\nu}\nu_{c,g}
    ,
\end{align}
we obtain
\begin{equation}
    \nu_{c,gg}  \sim
    - \frac{1}{t_0}
    \left(
    (V_g'+V_\nu'\nu_{c,g})\mathring{t}_0 + V_{gg}+V_{\nu g}\nu_{c,g} + (V_{g\nu}+V_{\nu\nu}\nu_{c,g}) \nu_{c,g}
     \right)
    .
\end{equation}
From \refeq{v-expansion},
\begin{equation}
    V_{gg} = -\frac{v_{gg}}{1+v} + \frac{v_g^2}{(1+v)^2} \sim -v_{gg} +O(t_0^2) \sim -M_{0,gg} t_0
    \sim -M_4^c t_0.
\end{equation}
Similarly, $V_{g\nu} \sim -M_3^c t_0$ and $V_{\nu\nu}\sim -M_2^ct_0$.
Therefore, since $\nu_{c,g} \sim -M_2^c$ by Lemma~\ref{lem:tangent}(ii),
and since $-\mathring{t}_0 \to B_3 = \frac{3b}{2\alpha}$ as $g \uparrow g_c$ by Proposition~\ref{prop:t0iii},
\begin{equation}
  \nu_{c,gg}   
  \sim
  \left( M_4^c -2M_2^cM_3^c + (M_2^c)^3 \right)
  + B_3 \frac{1}{t_0}
    (V_g'+V_\nu'\nu_{c,g}) .
\end{equation}

Let $f(g) = V_g'+V_\nu'\nu_{c,g}$.  Since
$f(g_c) = M_2^c + M_1^c(-M_2^c)=0$, we need the next order term.
Direct computation gives
\begin{align}
    V_g'+V_\nu'\nu_{c,g}
    & =
    \frac{1}{(1+v)^2}
    \left(
    -v_g'(1+v)+v'v_g - (-v'_\nu(1+v) + v'v_\nu)\frac{v_g}{v_\nu}
    \right).
\end{align}
By \refeq{v-expansion} and \refeq{Mderivs}, together with $M_0^c=M_1^c=1$
\begin{align}
    v & = M_0t + \frac 12 M_1 t^2 + \cdots \sim t + \frac 12 t^2 + \cdots,
    \\
    v' & = M_0+M_1t + \cdots \sim 1 + t + \cdots,
    \\
    v_g & = M_{0,g}t + \frac 12 M_{1,g}t^2 + \cdots \sim -M_2^c t - \frac 12 M_3^c t^2 + \cdots,
    \\
    v_\nu & = M_{0,\nu}t + \frac 12 M_{1,\nu}t^2 + \cdots \sim -  t - \frac 12 M_2^c t^2 + \cdots,
    \\
    v_g' & \sim -M_2^c-M_3^ct,
    \\
    v_\nu' & \sim -1-M_2^ct.
\end{align}
Therefore,
\begin{align}
    V_g'+V_\nu'\nu_{c,g}
    & \sim
    (M_2^c+M_3^ct)(1+t) -  M_2t
    -
    (
    (1+M_2^ct)(1+t)- t
    )
    \frac{M_2^c + \frac 12 M_3^c t}{1+ \frac 12 M_2^c t}
    \nnb & \sim
    M_2^c + M_3^ct - (1+M_2t)(M_2^c + \frac 12 M_3^c t)(1- \frac 12 M_2^c t)
    \nnb & \sim
    \frac 12 (M_3^c-(M_2^c)^2) t = \frac b2 t
    ,
\end{align}
and the proof is complete.
\end{proof}

\section*{Acknowledgements}
The work of GS was supported in part by {\small NSERC} of Canada.
We thank David Brydges for discussions concerning Section~\ref{sec:susy} and for showing us
the proof of Proposition~\ref{prop:Vint} via Lemma~\ref{lem:Vsusy}.
We are grateful to Anthony Peirce  and Kay MacDonald for assistance with {\small MATLAB} programming,
and to an anonymous referee for helpful advice.


\end{document}